\documentclass[iop,usenatbib,useAMS,numberedappendix,tighten,apj]{emulateapj}
\bibpunct{(}{)}{;}{a}{}{,}
\usepackage{enumerate}

\usepackage[T1]{fontenc}
\usepackage[latin1]{inputenc}
\usepackage{graphicx}
\usepackage{amssymb}
\usepackage{times}
\usepackage{amsmath}
\usepackage{rotating}

\newcommand{\tauv}{\hbox{$\hat{\tau}_{V}$}}

\newcommand{\ldust}{\hbox{$L_{\mathrm{dust}}$}}

\newcommand{\fmu}{\hbox{$f_\mu$}}

\newcommand{\mic}{\hbox{$\mu$m}}
\newcommand{\lsun}{\hbox{$L_\odot$}}
\newcommand{\msun}{\hbox{$M_\odot$}}
\newcommand{\mdust}{\hbox{$M_\mathrm{dust}$}}
\newcommand{\lfir}{\hbox{$L_\mathrm{FIR}$}}
\newcommand{\lir}{\hbox{$L_\mathrm{IR}$}}

\newcommand{\lpco}{\hbox{$L^{\prime}_\mathrm{CO}$}}
\newcommand{\cii}{\hbox{[CII]}}
\newcommand{\lcii}{\hbox{$L_\mathrm{[CII]}$}}

\def\lesssim{\mathrel{\hbox{\rlap{\hbox{\lower5pt\hbox{$\sim$}}}\hbox{$<$}}}}
\def\gtrsim{\mathrel{\hbox{\rlap{\hbox{\lower5pt\hbox{$\sim$}}}\hbox{$>$}}}}

\shorttitle{Empirical predictions for (sub-)mm deep fields}
\shortauthors{E. da Cunha et al.}

\begin{document}

%% LaTeX will automatically break titles if they run longer than
%% one line. However, you may use \\ to force a line break if
%% you desire.

\title{Empirical predictions for (sub-)millimeter line and continuum deep fields}

%% Use \author, \affil, and the \and command to format
%% author and affiliation information.
%% Note that \email has replaced the old \authoremail command
%% from AASTeX v4.0. You can use \email to mark an email address
%% anywhere in the paper, not just in the front matter.
%% As in the title, use \\ to force line breaks.

\author{Elisabete da Cunha\altaffilmark{1}, Fabian Walter\altaffilmark{1}, Roberto Decarli\altaffilmark{1}, Frank Bertoldi\altaffilmark{2}, Chris Carilli\altaffilmark{3}, Emanuele Daddi\altaffilmark{4}, \\
David Elbaz\altaffilmark{4}, Rob Ivison\altaffilmark{5,6}, Roberto Maiolino\altaffilmark{7,8}, Dominik Riechers\altaffilmark{9,10}, Hans-Walter Rix\altaffilmark{1}, \\
Mark Sargent\altaffilmark{4}, Ian Smail\altaffilmark{11}, Axel Weiss\altaffilmark{12}}
\affil{$^1$Max-Planck-Institut f\"ur Astronomie, K\"onigstuhl 17, 69117 Heidelberg, Germany}
\affil{$^2$Argelander Institute for Astronomy, University of Bonn, Auf dem H\"ugel 71, 53121 Bonn, Germany}
\affil{$^3$National Radio Astronomy Observatory, Pete V. Domenici Array Science Center, P.O. Box O, Socorro, NM, 87801, USA}
\affil{$^4$Laboratoire AIM, CEA/DSM-CNRS-Universit\'e Paris Diderot, Irfu/Service d'Astrophysique, CEA Saclay, \\
Orme des Merisiers, 91191 Gif-sur-Yvette Cedex, France}
\affil{$^5$ UK Astronomy Technology Centre, Royal Observatory, Blackford Hill, Edinburgh EH9 3HJ, United Kingdom}
\affil{$^6$ Institute for Astronomy, University of Edinburgh, Blackford Hill, Edinburgh EH9 3HJ, United Kingdom}
\affil{$^7$Cavendish Laboratory, University of Cambridge, 19 J.J. Thomson Avenue, Cambridge CB3 0HE, United Kingdom}
\affil{$^8$ Kavli Institute for Cosmology, University of Cambridge, Madingley Road, Cambridge CB3 OHA, United Kingdom}
\affil{$^{9}$Astronomy Department, California Institute of Technology, MC 249-17, 1200 East California Boulevard, Pasadena, CA 91125, USA}
\affil{$^{10}$ Department of Astronomy, Cornell University, Ithaca, NY 14853, USA}
\affil{$^{11}$Institute for Computational Cosmology, Durham University,  South Road,  Durham DH1 3LE,  United Kingdom}
\affil{$^{12}$Max-Planck-Institut f\"ur Radioastronomie, Auf dem H\"ugel 69, 53121 Bonn, Germany}

\email{cunha@mpia.de}

\begin{abstract}
Modern (sub-)millimeter/radio interferometers such as ALMA, JVLA and the PdBI successor NOEMA will enable us to measure the dust and molecular gas emission from galaxies that have luminosities lower than the Milky Way, out to high redshifts and with unprecedented spatial resolution and sensitivity. This will provide new constraints on the star formation properties and gas reservoir in galaxies throughout cosmic times through dedicated deep field campaigns targeting the CO/[CII] lines and dust continuum emission in the (sub-)millimeter regime. In this paper,
we present empirical predictions for such line and continuum deep fields. We base these predictions on
the deepest available optical/near-infrared ACS and NICMOS data on the Hubble Ultra Deep Field (over an
area of about 12 arcmin$^2$). Using a physically-motivated spectral energy distribution model, we fit the observed
optical/near-infrared emission of 13,099 galaxies with redshifts up to $z=5$, and obtain median likelihood estimates of their stellar mass, star formation rate, dust attenuation and dust luminosity. We combine the attenuated stellar spectra with a library of infrared emission models spanning a wide range of dust temperatures
to derive statistical constraints on the dust emission in the infrared and (sub-)millimeter which are consistent
with the observed optical/near-infrared emission in terms of energy balance. This allows us to estimate, for each
galaxy, the (sub-)millimeter continuum flux densities in several ALMA, PdBI/NOEMA
and JVLA bands. As a consistency check, we verify that the 850$\mu$m number counts and extragalactic background
light derived using our predictions are consistent with previous observations. Using empirical relations between the observed CO/[CII] line luminosities and the infrared
luminosity of star-forming galaxies, we infer the luminosity of the CO(1--0) and [CII] lines from the estimated infrared luminosity of each galaxy in our sample. We then predict the luminosities of higher CO transition lines CO(2--1) to CO(7--6) based on two extreme gas excitation scenarios: quiescent
(Milky Way) and starburst (M82). We use our predictions to discuss possible deep field strategies with ALMA. The predictions presented in this study will serve as a direct benchmark for future deep field campaigns in the (sub-)millimeter regime.

\end{abstract}

\keywords{
dust, extinction -- galaxies: evolution  -- galaxies: ISM -- galaxies: statistics -- submillimeter.
}

\section{Introduction} \label{sect:intro}

The last decade has seen impressive advances in our understanding of galaxy formation and evolution through galaxy surveys done (preferentially) at optical and infrared wavelengths. 
In particular, the history of star formation (the `Lilly--Madau' plot; e.g.~\citealt{Lilly1996,Madau1996,Hopkins2006}), and the build up of stellar mass as a function of galaxy
type and mass, have been well quantified to within 1 Gyr of the Big Bang. It has been shown that the comoving cosmic star formation rate density likely increases gradually
from $z \sim 6 - 10$, it peaks at $z \simeq 2$, and drops by more than an order of magnitude from $z\simeq1$ to $z\simeq0$ (\citealt{Hopkins2006,Bouwens2010}).
The build-up of stellar mass follows this evolution, as does the temporal integral \citep{Bell2003b,Ilbert2010}. The redshift range $z \simeq 1 - 3$, during which roughly half of the stars in the Universe were formed, is referred to as the `epoch of galaxy assembly'. In summary, the star formation properties as well as the stellar masses of galaxies have been well delineated
through these optical and near--infrared deep field studies; almost all of our current knowledge is based on optical and near-infrared deep fields of the stars, star formation, and ionized gas
(but see, e.g.~\citealt{Smolcic2009}, \citealt{Karim2011}, for additional constraints based on radio continuum studies). E.g., Lyman Break Galaxy samples have revealed a major population of star--forming galaxies at z$\simeq$3 (e.g.~\citealt{Steidel2003}). Likewise, magnitude-selected samples (e.g.~\citealt{LeFevre2005,Lilly2007}) provide a census of the
star-forming population based on UV/optical flux rather than color.

A key measurement that is currently (mostly) unavailable is that of the presence of molecular gas, i.e. the dense ISM phase (`fuel') required for star formation to take place, which lies at the heart of the evolution of the comoving cosmic star formation rate density. In recent years, there have been significant efforts devoted to obtaining molecular gas measurements of individual galaxies, by performing  follow-up studies of galaxies that have been  pre-selected from optical/NIR deep surveys. To date, color-selection techniques (e.g., `BzK', `BMBX') have revealed significant samples of gas-rich, star forming galaxies at $z \simeq 1.5$ to 2.5 (with molecular gas masses M$_{\rm H_2}\simeq10^{10}-10^{11}$\,M$_\odot$, stellar masses M$_\ast \simeq10^{10}-10^{11}$~M$_\odot$, and star formation rates SFR~$\simeq 100$~M$_\odot$~yr$^{-1}$; \citealt{Daddi2008,Daddi2010,Daddi2010b,Genzel2010,Tacconi2010,Geach2011}). While very important in their own right, these studies (that focus on the detection of carbon monoxide, the main tracer for molecular gas at low and high redshift) remain fundamentally limited to galaxy populations that were pre-selected in the optical/near-infrared, i.e. potentially missing gas-dominated and/or obscured systems. 

From a theoretical/modeling perspective, \cite{Obreschkow2009a,Obreschkow2009,Obreschkow2011} provide simulations of the cosmic evolution of the molecular (and atomic) hydrogen in galaxies as a function of redshift, by building on the Millennium dark matter simulations \citep{Springel2005} in which they place `idealized model galaxies' at the centers of the dark matter halos which then evolve according to simple rules (`semi-analytical modeling', \citealt{Obreschkow2009a,Obreschkow2009,Obreschkow2011}). \cite{Power2010} and \cite{Lagos2011} also present models of the cosmic evolution of the atomic and molecular gas content in galaxies by applying different semi-analytical galaxy formation models to the Millennium simulation.

In this paper, we present empirical predictions of molecular line and continuum deep fields that are only now becoming possible thanks to the advent of observational facilities that dwarf previous capabilities, in particular the broad bandwidth and sensitivity afforded by the Atacama Large Millimeter Array (ALMA), the Jansky Very Large Array (JVLA, formerly known as EVLA) and the IRAM Plateau de Bure Interferometer (PdBI) successor, the Northern Extended Millimeter Array (NOEMA). Our predictions are based on the deepest available optical and near-infrared data available for the Hubble Ultra Deep Field (UDF), but, barring cosmic variance, should give a statistical representation of an arbitrarily chosen region on the sky. Basically, we use a sophisticated spectral energy distribution (SED) model combined with a Bayesian approach \citep{daCunha2008} to interpret the observed optical/near-infrared emission of the UDF galaxies in terms of their stellar content, star formation activity and dust attenuation, and obtain statistical constraints on their total dust luminosity which are consistent with the observed stellar emission in terms of energy balance (i.e. all the stellar radiation absorbed by dust in the rest-frame ultraviolet to near-infrared is re-emitted in the mid-infrared to millimeter range). The dust luminosity of each model is then re-distributed at infrared to millimeter wavelengths by combining the (dust-attenuated) stellar SED with a library of infrared dust emission models spanning a wide range of dust SED shapes (including different dust temperatures). This allows us to derive median-likelihood estimates of the (sub-)millimeter continuum flux densities of our galaxies in several ALMA, PdBI/NOEMA
and JVLA bands. Based on these continuum predictions, we calculate predicted line strengths in the various rotational transitions of carbon monoxide (CO) and ionized carbon ([CII]), two main tracers of the star-forming interstellar medium in galaxies. We note that, with this technique, we potentially miss, by definition, extremely dust obscured sources that are not included in our optical/near-infrared catalog. However, this should not have a great impact in our results since the main goal of this paper is to characterize the general galaxy population rather than the extreme, dust-enshrouded starbursts.

In Section~\ref{sect:data}, we describe the main optical/near-infrared photometric catalogue of the Hubble UDF in which we base our predictions, as well as additional data from optical, infrared and sub-millimeter surveys of the UDF/Extended Chandra Deep Field South area that we use to test our predictions.  In Section~\ref{sect:modelling}, we describe our spectral energy model and fitting method.
In Section~\ref{sect:fit_results}, we analyze in detail our predicted (sub-)millimeter properties for 13,099 galaxies in the UDF. We discuss the SED fitting outputs and derived physical properties of the galaxies in our sample in Sections~\ref{sect:sed_results} and \ref{sect:properties}, respectively. In Section~\ref{sect:reliability}, we discuss the reliability of our infrared luminosity and (sub-)millimeter continuum flux estimates from the observed optical/near-infrared SEDs, and in Sections~\ref{sect:number_counts} and \ref{sect:ebl} we perform consistency checks on the predicted continuum flux densities of our galaxies at 850\mic\ by comparing them with observed number counts and the extragalactic background light. In Section~\ref{sect:results_continuum}, we present the distribution of our predicted continuum flux densities for the whole sample in several (sub-)millimeter bands from 38 to 870~GHz, and in Section~\ref{sect:lines} we infer the CO and \cii\ line luminosities of the galaxies in our sample from the estimates of their infrared luminosities, based on simple empirical relations. Based on these (sub-)millimeter line and continuum predictions, we discuss future deep fields with ALMA in Section~\ref{sect:discussion}. We summarize our main conclusions in Section~\ref{conclusion}. Appendix~\ref{appendix_seds} contains a comparison of our results with what we would obtain assuming fixed SED shapes for the galaxies in our sample.

Throughout this paper, we use a concordance $\Lambda$CDM cosmology with $H_0=70$~km~s$^{-1}$~Mpc$^{-1}$, $\Omega_\Lambda=0.7$ and $\Omega_m=0.3$.

%%%%%%

\section{The data}
\label{sect:data}

As an example, we here base our predictions on the deepest available optical and infrared data on the Hubble Ultra Deep Field (UDF) that will be accessible with ALMA. We stress that this coherent database is the only reason for our choice and that, barring the issue of cosmic variance, we could have used any other field for our predictions as well. This means that our statistical predictions should also hold for a northern field that will be accessible from other telescopes (e.g. IRAM PdBI/NOEMA, JVLA).

\subsection{Input catalog: optical/near-infrared HST data}
\label{sect:hst_data}

\begin{figure}
\begin{center}
\includegraphics[width=0.45\textwidth]{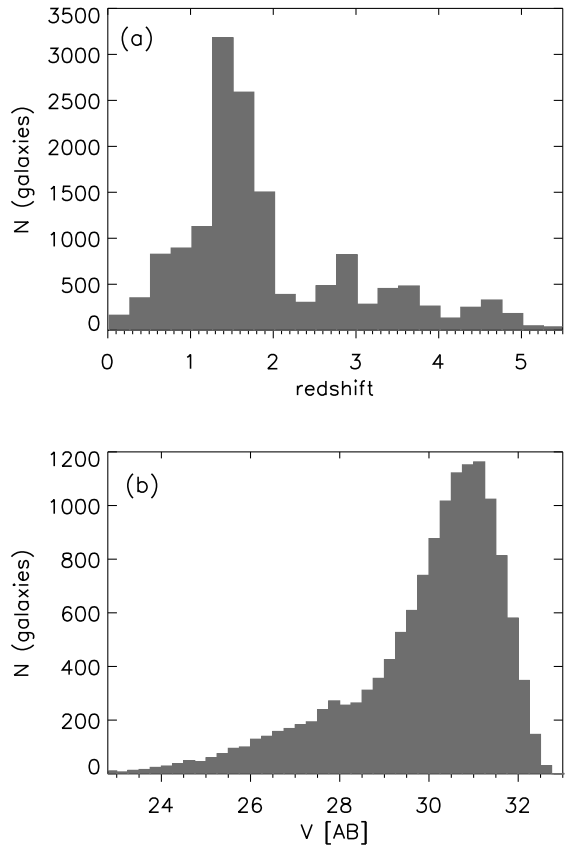}
\caption{Distribution of the photometric redshift (a) and V-band magnitude (b) of the optically-selected Hubble UDF galaxies from the \protect\cite{Coe2006} catalog.}
\label{redshift}
\end{center}
\end{figure}

We start by using the photometric catalog of the Hubble UDF (centered at R.A.$=03^\mathrm{h}32^\mathrm{m}39^\mathrm{s}.0$, Dec=$-27^\mathrm{\circ}47^\mathrm{\prime}29^\mathrm{\prime\prime}.1$) described in \cite{Coe2006}. This contains aperture-matched, PSF-corrected ACS $BVi^\prime z^\prime$ and NICMOS3 $JH$ photometry, as well as Bayesian photometric redshifts for all the detected sources, accurate to within $0.04(1+z_\mathrm{spec})$ \citep{Coe2006}. The full catalog contains $18,700$ sources, of which a large fraction (8,042) are detected at the $10\sigma$ level in at least one band. The $10\sigma$ limiting AB magnitudes in the $B$, $V$, $i^\prime$, $z^\prime$, $J$ and $H$ are 28.71, 29.13, 29.01, 28.43, 28.30 and 28.22, respectively. 
Following \cite{Coe2006}, to exclude contamination from stars, we exclude sources with $i$-band stellarity \texttt{stel} $\ge 0.8$ (about 6 per cent of the sample), leaving us with 17,532 galaxies, with photometric redshifts distributed as plotted in Fig.~\ref{redshift}(a). For reference, in Fig.~\ref{redshift}(b), we plot the distribution of the observed ACS $V$-band magnitude for 16,830 sources detected in that band. We note that the fact that the redshift distribution of our sources peaks at $z\simeq2$ and the sudden drop in sources fainter than 30 AB magnitudes in the $V$-band are due to incompleteness. This implies that our predictions may be missing high-redshift, dust-obscured galaxies that are too faint in the optical to be in our sample and may have large (sub-)millimeter fluxes. This is the case of LESSJ0333243.6-274644, the only sub-mm source detected in the UDF as part of the LESS survey (LABOCA observations of the Extended {\em Chandra} Deep Field South at 870~\mic; \citealt{Weiss2009}), which has no optical counterpart in our optical catalog. Our predictions are therefore lower limits for the possible number of detections at high redshift ($z>2$).

\subsection{Supporting data}
\label{sect:supporting_data}

We complement the photometry in the UDF catalog with additional photometry out to the far-infrared. We use 54 galaxies detected in the {\it Herschel}/SPIRE bands available in the publicly released HerMES survey (P.I. S. Oliver; \citealt{Oliver2010}) images in GOODS-South, for which we applied the same prior source extraction technique as in \cite{Elbaz2011} for the GOODS-{\it Herschel} SPIRE data in GOODS-North. {\it Herschel}/PACS images of the GOODS-South are available as part of the GOODS-{\it Herschel} program (P.I. D. Elbaz). For each of these 250\mic-selected galaxies, well-sampled spectral energy distributions from the ultraviolet to the far-infrared are available, including photometry in the following bands: $U$, $B$, $V$, $i$, $z$, $J$, $K$, {\it Spitzer}/IRAC 3.6, 4.5, 5.8 and 8.0~\mic, {\it Spizer}/IRS at 16~\mic, {\it Spitzer}/MIPS at 24~\mic, {\it Herschel}/PACS at 70, 100 and 160~\mic, and {\it Herschel}/SPIRE at 250, 350 and 500~\mic\ (\citealt{Elbaz2011}, \citealt{Magdis2011}). The redshifts of the galaxies in this sub-sample go from $z=0.140$ to $z=2.578$, with a median of value of $1.0$.
We use this sub-sample in Section~\ref{sect:reliability} to test the reliability of our predictions of the total infrared luminosity and (sub-)millimeter continuum fluxes from observed optical data as described in Section~\ref{sect:modelling}.

To test our predictions for a wider field and address the issue of cosmic variance in Section~\ref{sect:number_counts}, we use a wider-area catalog of the {\it Chandra} Deep Field South field which also includes the UDF but is about 10 times larger in area, the FIREWORKS catalog \citep{Wuyts2008}. The FIREWORKS catalog is a $K_s$-band selected galaxy catalog which contains multi-wavelength photometry of 6,308 galaxies from the $U$ band to the {\it Spitzer}-24\mic\ band, with a $5\sigma$ magnitude limit of 24.3~AB mag in the $K_s$ band (i.e. shallower than the UDF HST catalog described in Section~\ref{sect:hst_data}), over 138~arcmin$^2$.

\section{SED Modelling}
\label{sect:modelling}

In a next step we use the models described in \cite{daCunha2008} to fit the observed rest-frame optical to near-infrared spectral energy distributions of the galaxies from the photometric catalog described in Section~\ref{sect:hst_data}, and obtain statistical estimates of the infrared luminosities, (sub-)millimeter continuum flux densities and CO line luminosities for each galaxy in the sample.

\subsection{Ultraviolet to near-infrared emission}

We use the spectral synthesis model of \cite{Bruzual2003} to compute the integrated light
emitted by stars in galaxies for a wide range of metallicities (distributed between 0.02 and 2 times solar), ages (distributed between 0.1~Gyr and the age of the Universe at each redshift), and star formation histories (parameterized as exponentially declining with a wide range of timescales, and superimposed random bursts of star formation).
To account for the attenuation of starlight by dust, we describe the interstellar medium of galaxies using the two-component model of \cite{Charlot2000}: the ambient (diffuse) interstellar medium and the star-forming regions (birth clouds). This prescription accounts for the fact that stars are born in dense molecular clouds, which dissipate typically on a time-scale of $10^7$ yr. As a result, the non-ionizing continuum emission from young OB stars and line emission from their surrounding HII regions is absorbed by dust in these birth clouds and then in the ambient ISM, while the light emitted by stars older than $10^7$~yr propagates only through the diffuse ISM. The main free parameters of this model are the effective $V$-band optical depth seen by young stars in birth clouds, \tauv, and the fraction of \tauv\ contributed by dust in the diffuse ISM, $\mu$.
Using this model, we compute the attenuated stellar emission of galaxies and the total luminosity absorbed and re-radiated by dust in the birth clouds and the diffuse ISM.
We use the model library described in \cite{daCunha2008}, which
includes 50\,000 attenuated stellar spectra spanning a wide range of star formation histories, metallicities and dust optical depths.

\subsection{Infrared to millimetre emission}

In the context of the model described in \cite{daCunha2008}, the energy absorbed by dust is re-radiated at infrared wavelengths through four different components:
\begin{enumerate}[(i)]
\item the emission by polycyclic aromatic hydrocarbons (PAHs);
\item a hot mid-infrared continuum (with temperature in the range 130 -- 250~K);
\item emission by warm dust in thermal equilibrium (with temperature in the range 30 -- 60~K and dust emissivity index $\beta=1.5$);
\item emission by cold dust in thermal equilibrium (with temperature in the range 15 -- 25~K and dust emissivity index $\beta=2$). 
\end{enumerate}
\cite{daCunha2008} use a wide library of infrared emission spectra where the temperatures and relative contributions of each component to the total infrared luminosity span a wide range of values. The models in this library are then directly compared to infrared observations of galaxies to constrain each dust emission component. In this paper, since we hardly have any observational constraints on the infrared SED of the galaxies, we do not require such a wide range of models. Therefore, we will adopt a reduced set of dust emission models which reflect the range of infrared SED shapes of local, normal star-forming galaxies.

For this work, our goal is to obtain a range of possible infrared SEDs that are consistent with the observed optical/near-infrared emission in terms of their overall energy balance. Therefore, for simplicity, we fix the values of most free parameters controlling the shape of the infrared SEDs of galaxies to those of three representative infrared SEDs presented in \cite{daCunha2008}: a `standard' model (with equilibrium temperatures of the cold and warm dust components 22 and 48~K, respectively), a `hot' model (25 and 55~K) and a `cold' model (18 and 40~K); the relative contributions of the cold and warm dust components, as well as the PAHs and hot mid-infrared continuum are different for each model and are listed in Table~1 of \cite{daCunha2008}. These three models were calibrated using observed IRAS and ISO infrared fluxes of local star-forming galaxies and span the range of observed infrared colors for this low-redshift sample. The `standard' model reproduces the median infrared colors of local galaxies, the `hot' model is representative of the warmest observed infrared colors, and the `cold' model is representative of the cooler infrared colors. We build a simplified dust emission model library in which we fix the values of the dust temperatures and the contribution by PAHs, hot mid-infrared continuum and warm dust to the total luminosity of birth clouds, as well as the contribution of cold dust to the total dust luminosity of the diffuse ISM, to the values of these representative models, while leaving the fraction of total dust luminosity contributed by the diffuse ISM,  $\fmu=L_{\mathrm{dust}}^{\,\mathrm{ISM}}/L_{\mathrm{dust}}$, and the total dust luminosity, $L_{\mathrm{dust}}$ as free parameters.

\subsection{Radio continuum}

In addition to thermal dust emission, the (sub-)millimeter continuum emission of star-forming galaxies can have a non-negligible radio continuum emission component, especially at the lowest frequencies. This emission is mainly free-free emission from H~II regions and synchrotron radiation from relativistic electrons accelerated in supernova remnants (e.g.~\citealt{Condon1992}). Since the \cite{daCunha2008} models do not include radio emission, we add a radio continuum component to our model SEDs in order to account for this extra contribution to the (sub-)millimeter continuum. We use the simple prescription described in \cite{Dale2002}, which is based on the observed radio/far-infrared correlation.
The radio/far-infrared correlation (e.g. \citealt{Helou1985,Condon1992,Bell2003}) implies that the ratio between the observed far-infrared flux of a galaxy (between $42.5~\mic$ and $122.5~\mic$) and the radio flux density at $1.4$~GHz, $q$, is constant. We model the radio continuum in star-forming galaxies as a sum of two main components:
\begin{enumerate}[(i)]
\item a thermal component, consisting mainly of free-free emission from ionized gas, with spectral shape $f_\nu^\mathrm{\,th} \propto \nu^{-0.1}$;
\item a non-thermal component, consisting of synchrotron radiation, with spectral shape  $f_\nu^\mathrm{\,nth} \propto \nu^{-0.8}$.
\end{enumerate}
In normal star-forming galaxies, the contribution of the thermal (free-free) component to the radio continuum at 20~cm is $\sim 10\%$ \citep{Condon1992}. This allows us to fix the spectral shape of our radio continuum, which we normalize relative to the far-infrared flux of each model in our library using the value found by \cite{Yun2001}, $q=2.34$. This prescription is based on the assumption that the galaxies fall in the observed radio/far-infrared correlation, and has the advantage of not requiring any extra free parameters to estimate the radio continuum. We also assume that the radio/far-infrared correlation remains constant with redshift (e.g.~\citealt{Sargent2010}).

\subsection{SED fitting method}
\label{sect_method}

The libraries of attenuated stellar emission and dust emission are combined by associating models with similar values of
$\fmu=L_{\mathrm{dust}}^{\mathrm{\,ISM}}/L_{\mathrm{dust}}$
(the fraction of total dust luminosity contributed by the diffuse ISM), which are scaled to the same
total dust luminosity $L_{\mathrm{dust}}$. This ensures, for each model, the energy balance between the radiation absorbed 
and re-emitted by dust in the diffuse ISM and stellar birth clouds.

For each galaxy, we compare the observed optical fluxes in the ACS and NICMOS bands (Section~\ref{sect:data})
to the predicted fluxes for every model of the stochastic library described above, by computing the
$\chi^2$ goodness of fit for each model.
We then build the likelihood distribution of any given physical parameter
for the observed galaxy by weighting the
value of that parameter in each model by the probability $\exp(-\chi^2/2)$. Our final estimate of the parameter
is the median of the likelihood distribution, with an associated confidence
interval which is the 16th--84th percentile range of the distribution (this confidence interval is tighter for well-constrained parameters
and wider when the parameters are not well constrained by the available observations). In what follows, the
values of the physical properties of the galaxies mentioned refer to the median values
of the probability density distribution.
We also obtained the best-fit model SED
for each galaxy, which is the model that minimizes $\chi^2$. 

\section{Predicted (sub-)millimeter properties}
\label{sect:fit_results}

\subsection{SED fitting outputs}
\label{sect:sed_results}

\begin{figure*}
\centering
\includegraphics[width=0.75\textwidth]{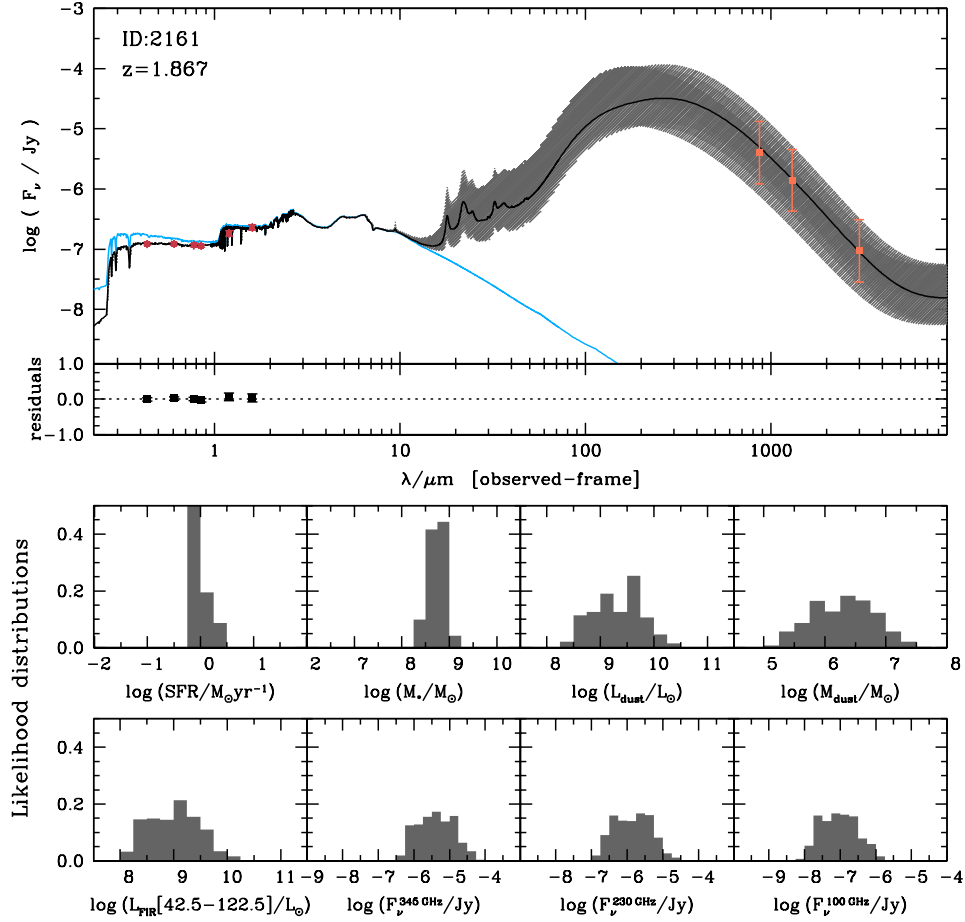}
\caption{Example of a fit to the observed optical spectral energy distribution of a UDF galaxy at $z=1.867$ (red points). The black line represents the best-fit model SED; the blue line represents the unattenuated (i.e. dust-free) stellar spectrum. The grey-shaded area represents the range of all infrared dust emission models in our model library that are consistent with the observed optical/near-infrared fluxes in terms of energy balance. The orange points represent the median of the probability density function (PDF) for the predicted fluxes at 345, 230 and 100~GHz (three arbitrary chosen (sub-)mm bands), and the associated error bars represent the confidence range, i.e. the 16th--84th percentile range of the PDF. The residuals of the fit are plotted in the panel at the bottom of the SED. The 8 bottom panels show the PDFs of several parameters: star formation rate; stellar mass; total dust luminosity; dust mass; far-infrared luminosity between 42.5 and 122.5~$\mu$m; and flux densities in the three randomly chosen bands at 345, 230 and 100~GHz.}
\label{example_fit}
\end{figure*}

We use the method described above to fit the observed photometry of each galaxy and produce likelihood distributions of their stellar mass, star formation rate, dust optical depths, infrared luminosities, dust masses, continuum and molecular line fluxes in the (sub-)millimeter range.

In Fig.~\ref{example_fit}, we show an example SED fit and the associated likelihood distributions of some physical parameters: the star formation rate (SFR), stellar mass ($M_\ast$), dust luminosity (\ldust), dust mass (\mdust), far-infrared luminosity (\lfir, defined as the integral of the infrared emission between 42.5 and 122.5~$\mu$m), and the continuum flux densities in a number of accessible (sub-)millimeter windows at 345, 230 and 100~GHz (specifically, ALMA bands 7, 6 and 3, respectively, and PdBI/NOEMA band 1, 3 and 4). The ultraviolet to near-infrared part of the SED is the stellar population model that best fits the data. The far-infrared and (sub-)millimeter part of the SED that is plotted corresponds to the model with the best fit $L_{\mathrm{dust}}$ and $\fmu$, but all the other parameters controlling the shape of the SED at these wavelengths are randomly drawn from the library of dust emission models. The grey shaded area shows the range of possible infrared SEDs allowed within the uncertainties in dust luminosity with different dust temperatures and contributions of PAHs, mid-IR continuum and dust in thermal equilibrium reflecting the diversity of possible dust emission models in our model library. The orange points show the (exemplary) median-likelihood estimates of the flux densities at 345, 230 and 100~GHz, where the error bars (16th -- 84th percentile range) reflect all the different combinations of infrared models that are consistent with the observed optical data.

We impose a minimum of three photometric bands to define the SED, and we discard galaxies with $z>5$ and galaxies for which the fit residuals are larger than $2\sigma$ (where $\sigma$ is the uncertainty associated with the flux measurement) in each band. These criteria allow us to discard the most unreliable SED fits: at very high redshift, our model becomes uncertain and the observations sample only the far-UV emission, making it very difficult to constrain the SED; also, galaxies with very high residuals in a given band may indicate problems with the photometry or wrong photometric redshift, or the presence of an AGN (which is not included in our models). Our final sample used in the remainder of this paper consists of $13,099$ sources with $z\le5$.

\subsection{Derived physical properties}
\label{sect:properties}

\begin{figure*}
\begin{center}
\includegraphics[width=\textwidth]{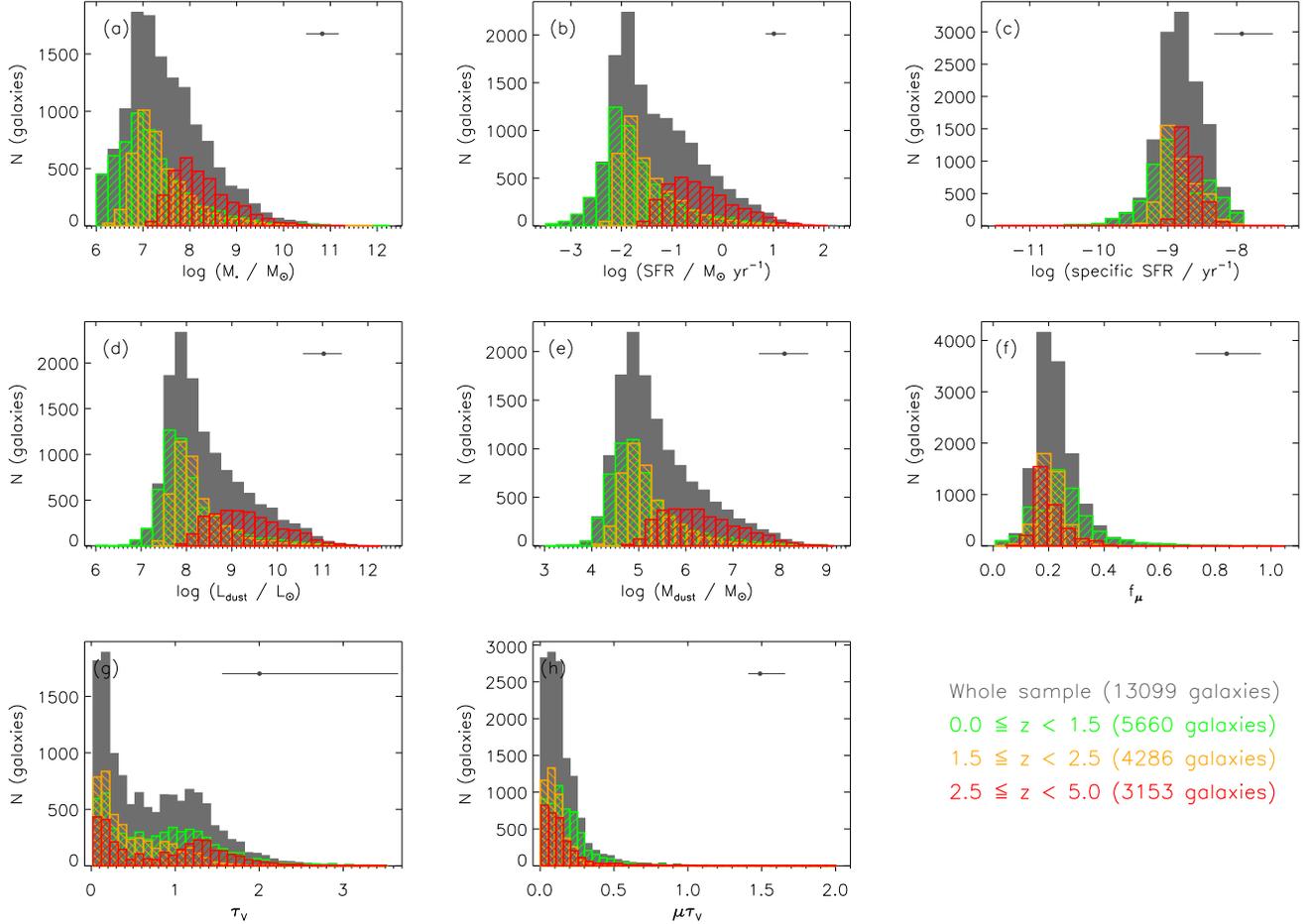}
\caption{Distributions of the median-likelihood parameters derived from SED fitting for the whole Hubble UDF galaxy sample. (a) stellar mass; (b) star formation rate averaged over the last 100~Myr; (c) specific star formation rate, defined as the star formation rate divided by stellar mass;
(d) total dust luminosity; (e) dust mass; (f) fraction of total dust luminosity contributed by the diffuse ISM, \fmu;
(g) effective $V$-band optical depth seen by stars younger than 10~Myr in birth clouds, \tauv; (h) effective $V$-band optical depth seen by stars older than 10~Myr in the diffuse ISM.
The grey histograms represent the whole sample, and the colored histograms represent the distribution of parameters of galaxies divided in three redshift bins: green: $z<1.5$; yellow: $1.5 \le z < 2.5$; red: $z > 2.5$. The error bars in the top right-hand corner of each plot represent the
median confidence range for each parameter. We note that the sharp drop towards lower values of stellar mass, SFR and dust mass/luminosity is due to incompleteness of the photometric catalog towards fainter flux levels (see Fig.~\ref{redshift}).}\label{hist_parameters}
\end{center}
\end{figure*}

Stellar masses are constrained to within $\pm 0.35$~dex, which reflects uncertainties due to the fact that, for most of our galaxies, observations do not include the rest-frame near-infrared, where the light is dominated by low-mass stars, which constitute the bulk of the stellar mass in galaxies. However, we show in the next section (Section~\ref{sect:reliability}) that this does not cause any systematic effects on the stellar mass estimates. The star formation rates are constrained to within typically $\pm 0.2$~dex, due to the fact that the observed SEDs sample the emission by young stars in the rest-frame ultraviolet. The dust luminosities are more uncertain (confidence ranges are typically $\pm 0.45$~dex), as expected due to the lack of infrared observations for our sample. The dust luminosity is estimated by our model by calculating the total energy absorbed by dust, taking into account the light emitted by stars and the attenuation by dust. Therefore, by construction, our dust luminosities are consistent with all stellar and dust attenuation parameters (SFR, $M_\ast$, $\mu$, \tauv, \fmu), from an energy balance perspective (as described in \citealt{daCunha2008} and in Section~\ref{sect:modelling}). Even though we have significant uncertainties in the dust luminosity estimates, as expected from our sparse SED sampling, we are still able to predict \ldust\ well within an order of magnitude (we analyze possible systematic effects in Section~\ref{sect:reliability}). The dust masses are also estimated by using all the dust emission model templates that are consistent with the statistical estimates on dust luminosities. The confidence range for \mdust\ is very large ($\pm 0.55$~dex), and reflects not only the uncertainty in \ldust, but also the large uncertainty in infrared SED shapes and dust temperatures. These dust masses are merely indicative of the range of dust masses that are consistent with the observed SEDs in terms of energy balance, and taking into account a range of possible dust temperatures, 18--25~K for the cold dust (with $\beta=2$), and 40--55~K for the warm dust (with $\beta=1.5$).

The distributions of physical parameters inferred from our SED fits (Fig.~\ref{hist_parameters}) show that the bulk of our galaxies are low-mass, low star formation rate and low dust attenuation (\tauv\ and $\mu\tauv$) sources, i.e. blue star-forming galaxies (consistent with the finding of a large population of faint blue galaxies in the UDF described in \citealt{Coe2006}).
As expected, galaxies in the highest redshift bin, $z \ge 2.5$, have typically higher stellar masses and star formation rates, because only the brightest galaxies are detected. The median dust luminosity of the sources, \ldust\ increases from $\log(\ldust/L_\odot) \simeq 8.0$ in the lowest redshift bins to $\log(\ldust/L_\odot) \simeq 9.2$ at $z\ge2.5$. Figs.\ref{hist_parameters}(g) and \ref{hist_parameters}(h) show that this is not necessarily due to an increase in the dust optical depth of galaxies in the highest redshift bin, but rather to the fact that the galaxies detected have larger stellar masses and star formation rates, as shown in Figs.\ref{hist_parameters}(a) and \ref{hist_parameters}(b).

\subsection{Reliability of infrared luminosity estimates}
\label{sect:reliability}

\begin{figure*}
\begin{center}
\includegraphics[width=0.975\textwidth]{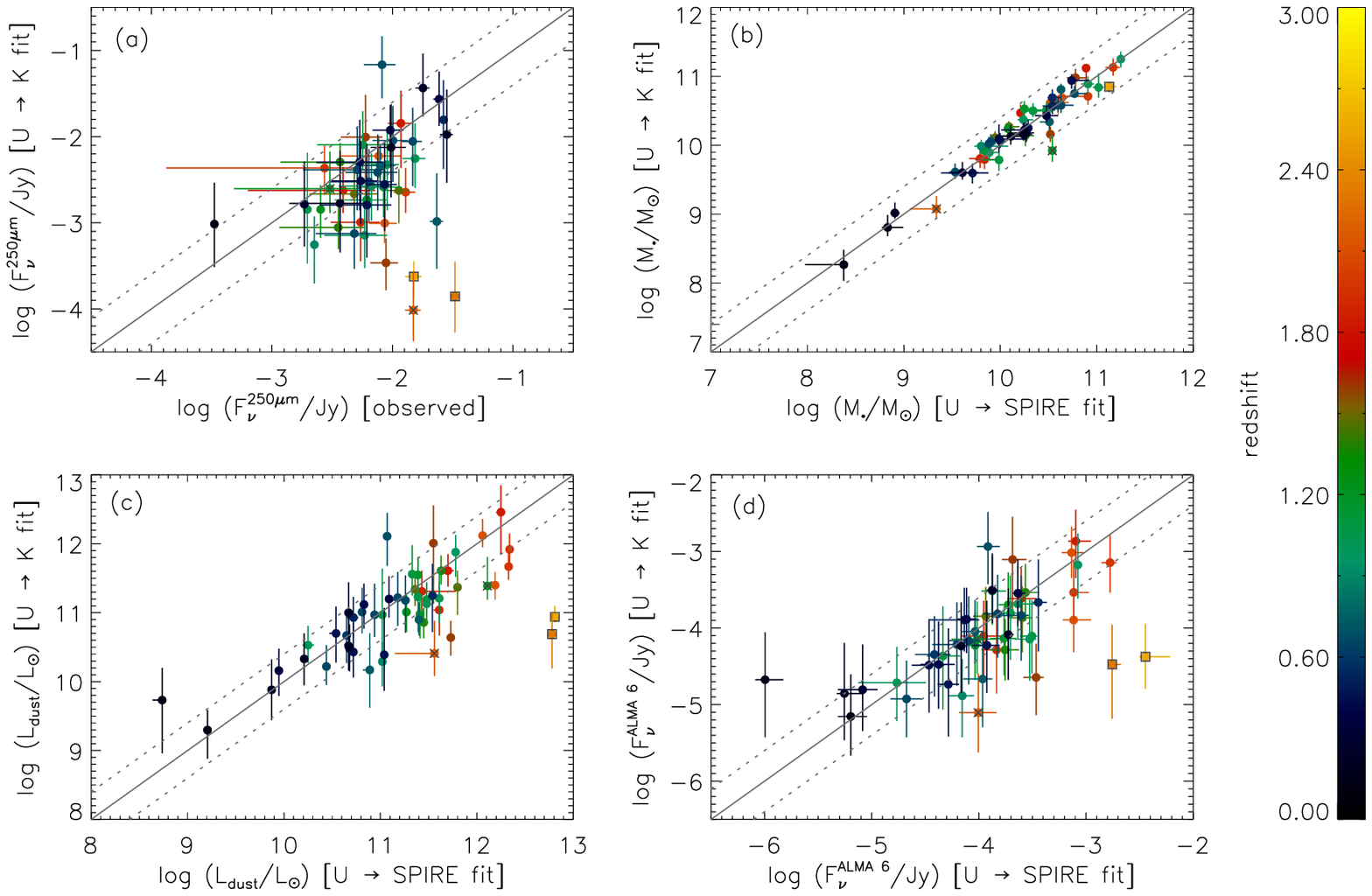}
\caption{(a) Comparison between the observed {\it Herschel}/SPIRE 250-\mic\ flux (x-axis) and our Bayesian median-likelihood estimate of the 250-\mic\ flux of each galaxy based on SED fits from the $U$-band to the $K$-band (y-axis). The other three panels show the comparison between our Bayesian median-likelihood estimates of GOODS/{\it Herschel} galaxy parameters obtained when fitting the full SED from the U-band to the longest available SPIRE band (x-axis), and when fitting only the SED from the $U$-band to the $K$-band (y-axis): (b) stellar mass; (c) total dust luminosity; (d) continuum flux in the ALMA band 6 (230~GHz). Each galaxy is color-coded according to redshift. The error bars show the 16th -- 84th percentile range of the likelihood distributions. In all panels, the grey solid line is the identity line, and the dotted lines show a $\pm 0.5$~dex offset for reference. The two points marked with crosses are galaxies that show a significant AGN contribution in the infrared; the two points marked with squares are galaxies which show a ULIRG-like SED, i.e. they seem to be very optically thick (given their high intrinsic infrared-to-optical emission ratios). Our SED modelling may not be reliable for these four galaxies, but overall we find a good agreement between the estimates derived from fitting the full SED and those from fitting the SED only up to the $K$-band.}
\label{compare_goodsh}
\end{center}
\end{figure*}

It is important to test how well we can predict the total infrared luminosity of the galaxies in our sample from their observed rest-frame UV/optical spectral energy distributions.
To do so, we use the sample of 54 UDF galaxies detected in the {\it Herschel}/PACS and {\it Herschel}/SPIRE bands as part of the GOODS-{\it Herschel} program described in Section~\ref{sect:supporting_data}. For each of these galaxies, we have well-sampled SEDs from the ultraviolet to the far-infrared, which allows us to test our SED extrapolations. We fit the observed (more complete) SEDs of this subsample of galaxies using the same method as described in Section~\ref{sect_method}. First, we include the full observed ultraviolet to far-infrared observations in the SED fits, not only to check that our model can reproduce consistently the SEDs of these galaxies, but also to get the best possible estimates of the stellar masses, star formation rates, dust luminosities and continuum (sub-)mm fluxes for these sources. Then, we re-fit the SEDs using only observations between the $U$ band and the $K$ band, to mimic the set of observations available for the majority of galaxies discussed above.

In Fig.~\ref{compare_goodsh}(a), we compare the median-likelihood of the {\it Herschel}/SPIRE 250-\mic\ flux derived from the fits from the $U$ to $K$ band with the actual observed 250-\mic\ flux for each galaxy.
We find a small systematic offset of 0.25~dex between the observations and our estimates, in the sense that we tend to underestimate the 250-\mic\ flux of the galaxies on average with our $U$-to-$K$-band fits. However, for most of the galaxies, the observed value is still within the confidence range derived from our fits. This effect is likely to be less important for the total dust luminosity and (sub-)mm fluxes, as these do not depend as strongly on the exact location of the peak of the infrared SED (i.e. the actual dust temperature spanned by the models).
In the next three panels of Fig.~\ref{compare_goodsh}, we compare the median-likelihoods of the stellar masses, dust luminosities, and continuum ALMA Band 6 fluxes obtained with the two sets of SED fits ($U$-to-$K$-band fit vs. $U$-to-SPIRE-fit). Fig.~\ref{compare_goodsh}(b) shows that the stellar masses agree remarkably well between the two fits (with a dispersion around the identity line of $0.13$~dex). Not surprisingly, the total dust luminosities and predicted ALMA band 6 fluxes do not agree as well, as shown in Figs.~\ref{compare_goodsh}(c) and (d). The inclusion of infrared data in the fits helps constrain these properties better, as shown by the significantly reduced confidence ranges. In the case of \ldust, this happens because the infrared data allow us to constrain the bolometric dust luminosity by fitting the dust emission itself (as opposed to constraining \ldust\ from the attenuated spectrum alone); the constraints on SFR are also tightened because we can account for dust-obscured star formation rate more accurately. The different far-infrared fluxes obtained with PACS and SPIRE help constrain the shape of the dust SED, namely the dust temperatures and relative contributions of the warm and cold dust components to the total infrared emission. This helps obtaining tighter constraints on the (sub-)mm continuum fluxes (namely the ALMA Band 6 flux shown as an example in Fig.~\ref{compare_goodsh}(d)).

Even though the inclusion of infrared data helps constraining parameters such as the star formation rate, dust luminosity and ALMA continuum fluxes, the median likelihood estimates of these parameters when {\em excluding} the infrared data agree very well with the estimates derived from the full SED fits, even if, as expected, the associated confidence ranges are larger. We find very small offsets between the averages of the median-likelihood estimates derived from the two fits: $0.02$~dex for SFR, $0.08$~dex for \ldust, and $0.07$~dex for the predicted ALMA Band 6 continuum flux (in the sense of the $U$-to-$K$-band fit slightly underestimating the parameters), with a dispersion of $\simeq 0.40$~dex for all cases. These very small systematic offsets are well within our fit confidence ranges, and show that our approach to predict infrared luminosities and (sub-)mm continuum fluxes from modelling UV/optical SEDs is reliable.
We note however that the difference between 250\mic\ fluxes and total dust luminosities derived from the fits to the UV/optical data only and those measured with {\it Herschel} correlates
with the dust optical depth in the galaxies. We tend to underestimate the (sub-)mm fluxes/dust luminosity when using only the $U$-to-$K$-band fits for galaxies with the highest dust attenuations (which translate into high infrared-to-optical ratios). This is due to the fact that our dust attenuation prior (Section~\ref{sect:modelling}; \citealt{daCunha2008}) leads to an underestimation of the optical depth for extremely dust-enshrouded, starburst-like sources (such as local ULIRGs or high-redshift SMGs; see \citealt{daCunha2010b}). While these galaxies can be a negligible fraction of our sample in number (e.g.~\citealt{Rodighiero2011,Sargent2012}), they can dominate the bright (sub-)mm counts. This is the case for the two GOODS/{\it Herschel} sources in our sample with the highest redshift, which are marked in Fig.~\ref{compare_goodsh} with squares. Due to the flux limit of this sample, at the highest redshifts ($z\simeq2$), only very dust-obscured ULIRG-type galaxies were selected. For this particular type of galaxies, the SED models used in Section~\ref{sect:modelling} become limited. However, we expect this kind of galaxies to be rare in our optically-selected sample of the UDF, and therefore we do not expect this limitation to greatly affect our results. We also note that our optically-selected catalogue is also likely to miss completely optically-obscured galaxies (e.g. HDF850.1, \citealt{Walter2012}; GN10, \citealt{Daddi2009b}; GN20, \citealt{Daddi2009a}).
The very good agreement between parameters derived from fits to the UV/optical SED versus parameters derived from fits to the full SED is consistent with previous results showing that the star formation properties of normal star-forming galaxies up to $z\simeq 2$ can be reliably derived from UV/optical observations alone (e.g. \citealt{Daddi2005,Daddi2007,Reddy2006}). This implies that the ISM of these galaxies is not heavily optically-thick, and we can apply our energy balance technique to interpret the SEDs of most normal star-forming `main sequence' galaxies.
 
\subsection{Number counts}
\label{sect:number_counts}

\begin{figure*}
\begin{center}
\includegraphics[width=0.65\textwidth]{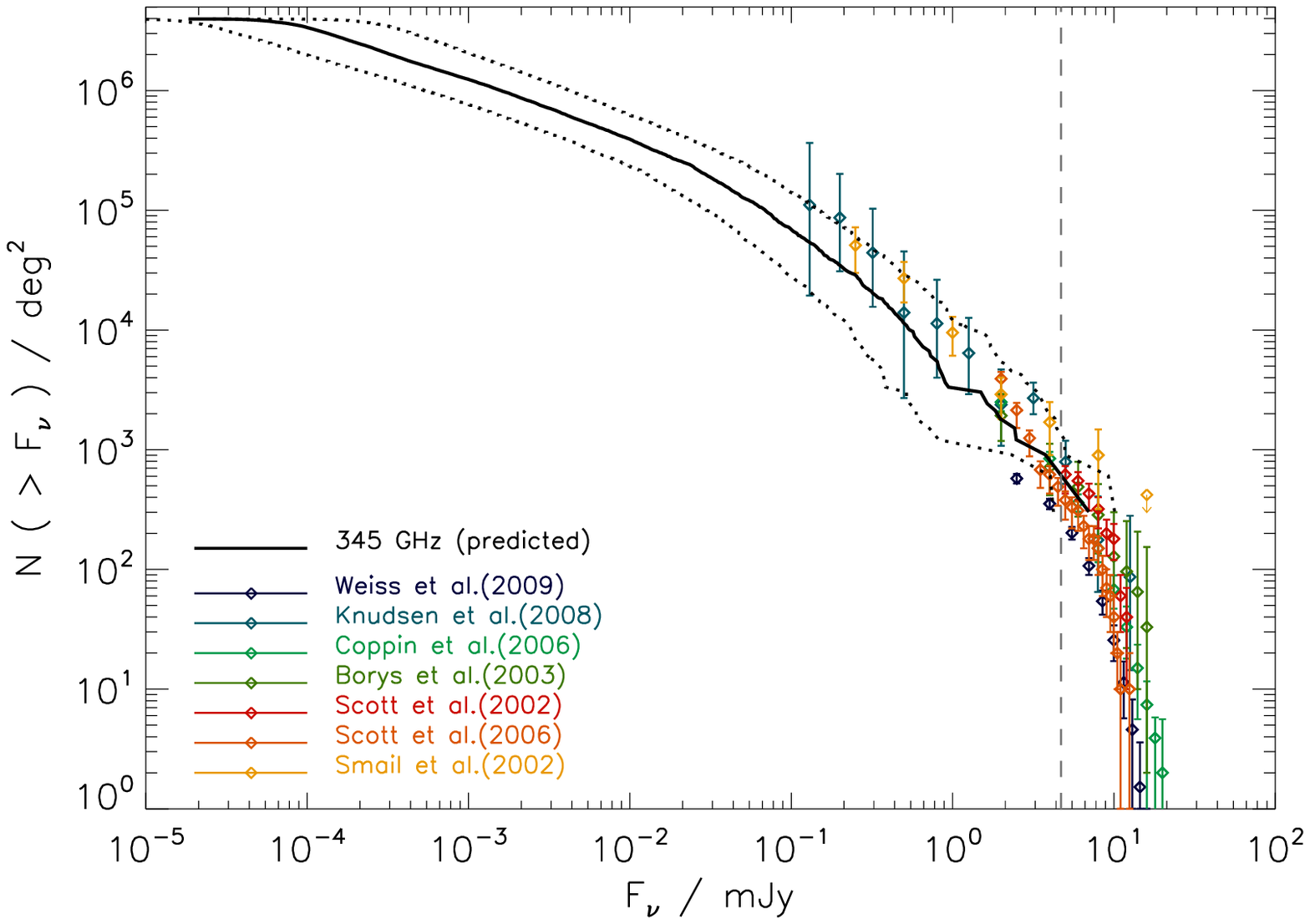}
\caption{Predicted cumulative number counts in Band 7 at 345~GHz (black solid line), with confidence range estimated using the upper and lower flux limits given by the confidence range for each galaxy (dotted lines). The colored points show observed values at $\sim$850\mic\ from different studies (see figure legend for references). The vertical line shows the flux limit of the LESS survey catalog \protect\citep{Weiss2009}, which includes the UDF.}
\label{cumulative}
\end{center}
\end{figure*}

As a consistency check, we now compare our continuum flux density predictions with previously obtained number counts at (sub-)millimeter wavelengths.

Number counts at 850\mic\ have been obtained using the SCUBA bolometer array on the JCMT and LABOCA on APEX by a number of groups over the last decade (e.g.~\citealt{Scott2002,Smail2002, Borys2003, Coppin2006,Scott2006,Knudsen2008,Weiss2009}). These counts can be directly compared to our predictions at 345~GHz (ALMA band 7, PdBI/NOEMA band 4) since this band probes roughly the same wavelength.
In Fig.~\ref{cumulative}, we compare our predicted cumulative number counts at 345~GHz with these previous observations. Our predicted number count range to first order agrees with the observed number counts, but we do not reach the higher fluxes probed by sub-mm observations. The lack of the brightest sources is due to two reasons. First, the field on which we base our predictions is very small (the size of the UDF is only $3.3\times10^{-3}$~deg$^2$), i.e. we do not expect the presence of a significant population of bright sources in the field. Second, we are working with an optically-detected sample, and the bright sub-mm counts are dominated by optically thick sources which are likely not detected in the optical. For example, LESSJ0333243.6-274644, the only sub-mm source detected in the Hubble UDF as part of the LESS survey (LABOCA observations of the Extended {\it Chandra} Deep Field South), with a flux density of 6.4~mJy at 870\mic\ \citep{Weiss2009}, has no optical counterpart in our catalog, presumably because it is an optically-thick sub-mm galaxy (SMG; cf. \citealt{Dunlop2011}). We find that the number counts at fluxes $\gtrsim 0.5$~mJy are dominated by galaxies with $\mu\tauv > 1$, i.e. where the ISM is optically-thick on average.

As an additional test on our number count predictions, we turn to a wider field covered by the LESS survey. To do so, we expand our analysis to the FIREWORKS photometric catalog \citep{Wuyts2008} on the CDF-S, described in Section~\ref{sect:supporting_data}. FIREWORKS covers an area that is about 10 times the area of the UDF and about 10 times smaller than the full E-CDFS. The photometric catalog is much shallower than that for the UDF. For the area covered by FIREWORKS, we estimate between 6 and 23 sources to have 870-\mic\ fluxes above 4.7 mJy (the flux limit of the LESS catalog). For comparison, \cite{Weiss2009} find 10 sources over the same area. Our prediction is thus broadly consistent with the LESS measurements.

\subsection{Extragalactic background light}
\label{sect:ebl}

We can also compare our predictions with measurements of the integrated extragalactic background light in the sub-mm. Using our median-likelihood estimate of the 345~GHz flux density for each galaxy in the Hubble UDF, we obtain an integrated continuum 870-\mic\ flux density of 45.6~Jy deg$^{-2}$. If we add the contribution from LESSJ0333243.6-274644 (which is not part of our sample but is detected in the LESS survey with a 6.4~mJy flux), we obtain an EBL value of 47.5~Jy~deg$^{-2}$. This value is fully 
consistent with measurements of the extragalactic background light from COBE observations  $44 \pm 15$~Jy~deg$^{-2}$ \citep{Puget1996,Fixsen1998}. We note that the use of fixed spectral energy distribution templates to derive the (sub-)millimeter continuum fluxes of the galaxies would lead to EBL values that are inconsistent with the observed value (see Appendix~\ref{appendix_seds}). In Table~\ref{ebl_tab}, we list our estimates of the EBL in different (sub-)millimeter bands.

The estimated EBL at 870~\mic\ using the FIREWORKS catalog over the CDF-S area is 36.3~Jy~deg$^{-2}$, broadly consistent with our estimate based on the UDF area only (Table~\ref{ebl_tab}). This is lower than the COBE observed value quoted above, but it is still consistent with the observations, since the FIREWORKS catalog does not reach very deep, and therefore it is likely to miss the large number of faint galaxies that make up for a significant fraction of the extragalactic background light.

\begin{deluxetable}{lcc}
\tablecaption{Predicted extragalactic background light.}
\tablewidth{0pt}
\tablecolumns{3}
\tablehead{
\colhead{Frequency} &
\colhead{Observatory} &
\colhead{EBL} \\
 \colhead{/ GHz} & 
 \colhead{}& 
 \colhead{/ Jy deg$^{-2}$}
}
\startdata
 38 & ALMA 1, JVLA Ka & 0.10 \\
80 & ALMA 2 & 0.74 \\
100 & ALMA 3, PdBI/NOEMA 1 &1.53 \\
144 & ALMA 4, PdBI/NOEMA 2 & 4.48 \\
230 & ALMA 6, PdBI/NOEMA 3 & 18.2 \\
345 & ALMA 7, PdBI/NOEMA 4 & 45.6 \\
430 & ALMA 8 & 63.1 \\
660 & ALMA 9 & 123 \\
870 & ALMA 10 & 146 \\
\enddata
\label{ebl_tab}
\tablecomments{Estimates of the extragalactic background light at different frequencies using our flux predictions (area of the UDF field is 0.0033~deg$^2$) for the 13,099 galaxies in our sample.}
\end{deluxetable}

\subsection{Continuum flux density predictions}
\label{sect:results_continuum}

In this section, we present our general predictions for the continuum flux densities of our galaxies. In Table~\ref{cont_predictions_table}, we make our continuum predictions in all relevant (sub-)mm bands available for all the galaxies in our sample.

\begin{figure*}
\begin{minipage}{0.5\linewidth}
\centering
\includegraphics[trim = 0.25cm 0cm 0.05cm  0.1cm, clip, width=0.79\textwidth]{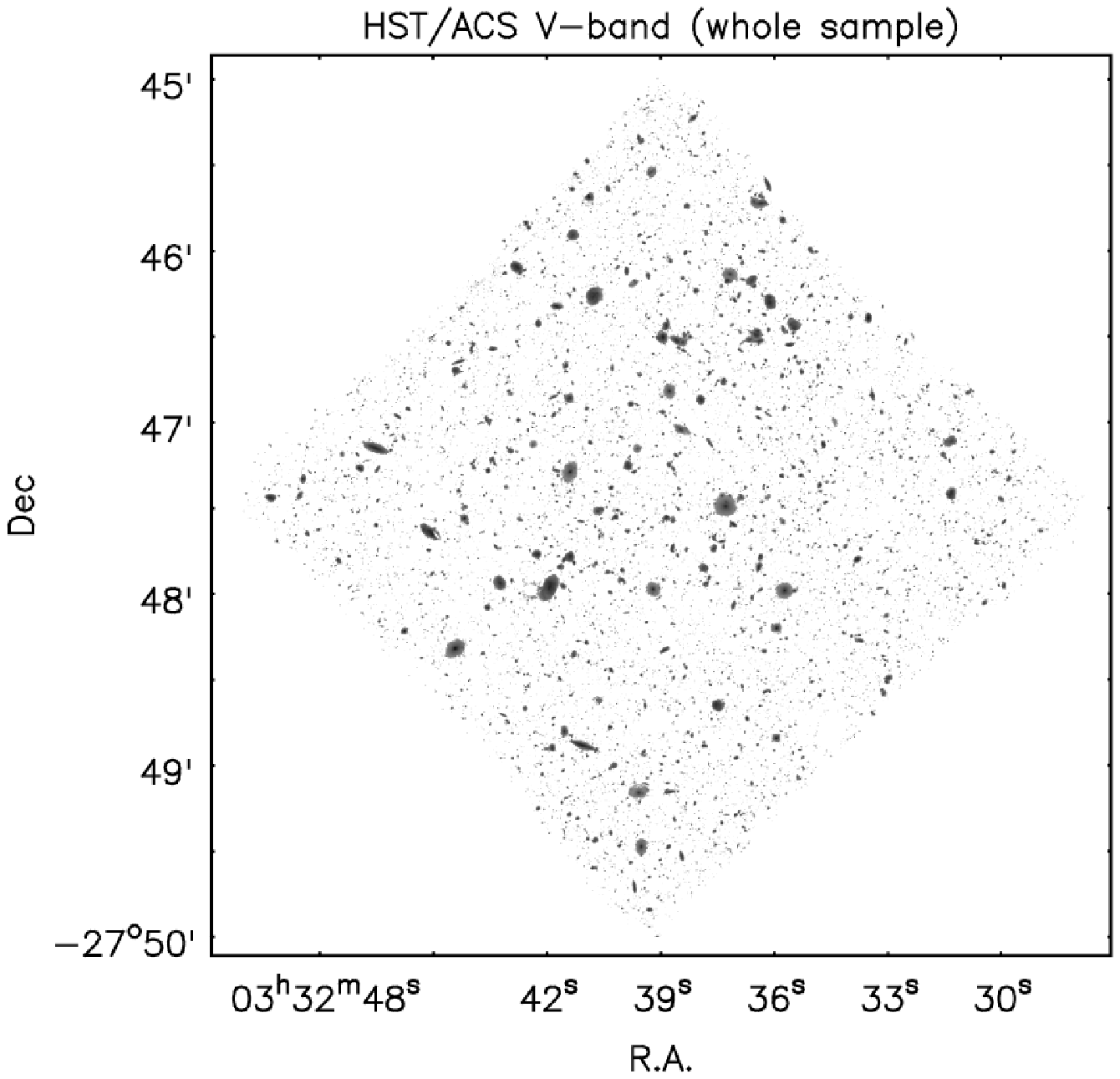}
\end{minipage}
\begin{minipage}{0.5\linewidth}
\centering
\includegraphics[trim = 0.25cm 0cm 0.1cm  0.1cm, clip,width=\textwidth]{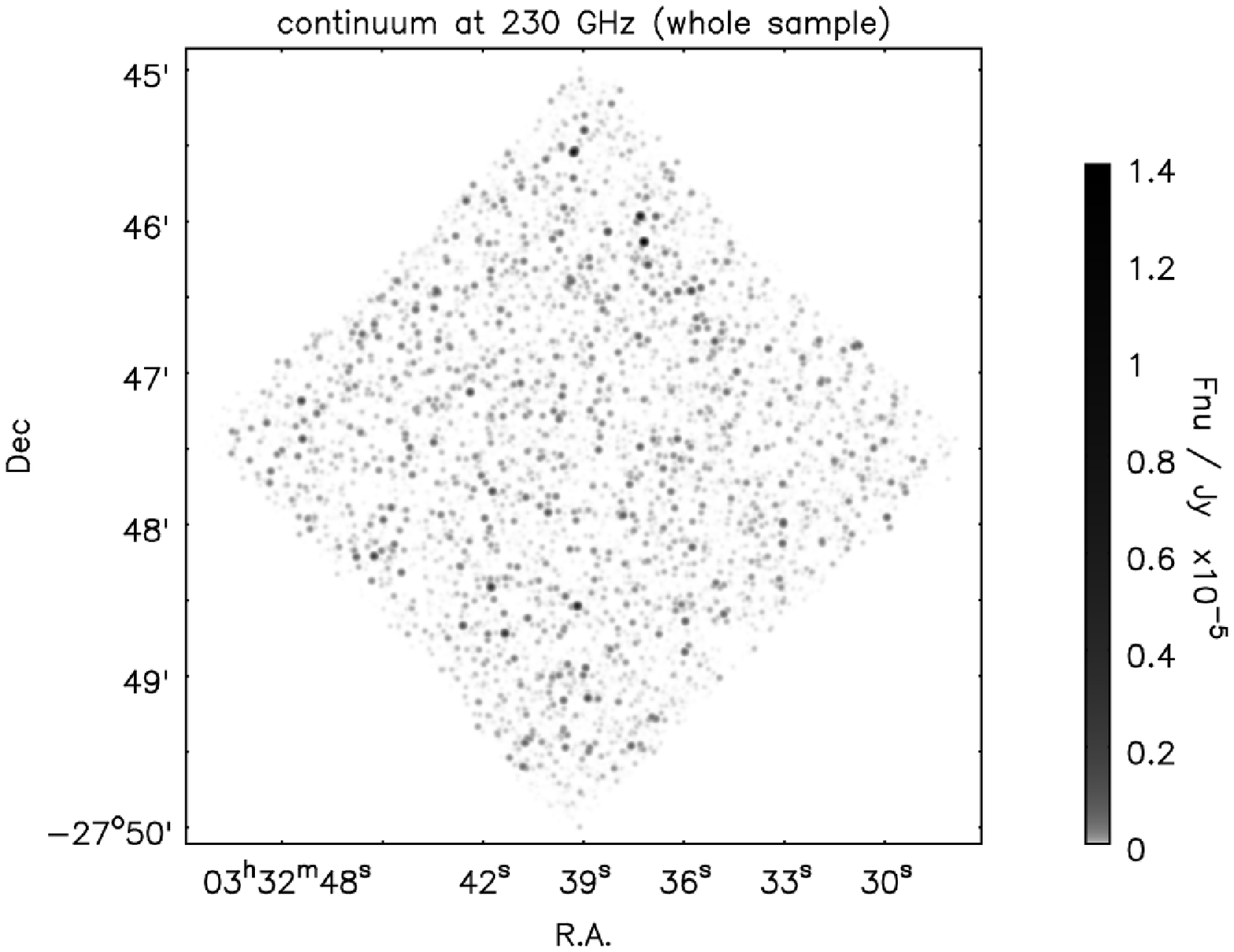}
\end{minipage}
\begin{minipage}{0.5\linewidth}
\centering
\includegraphics[trim = 0.25cm 0cm 0.05cm  0.1cm, clip,width=0.79\textwidth]{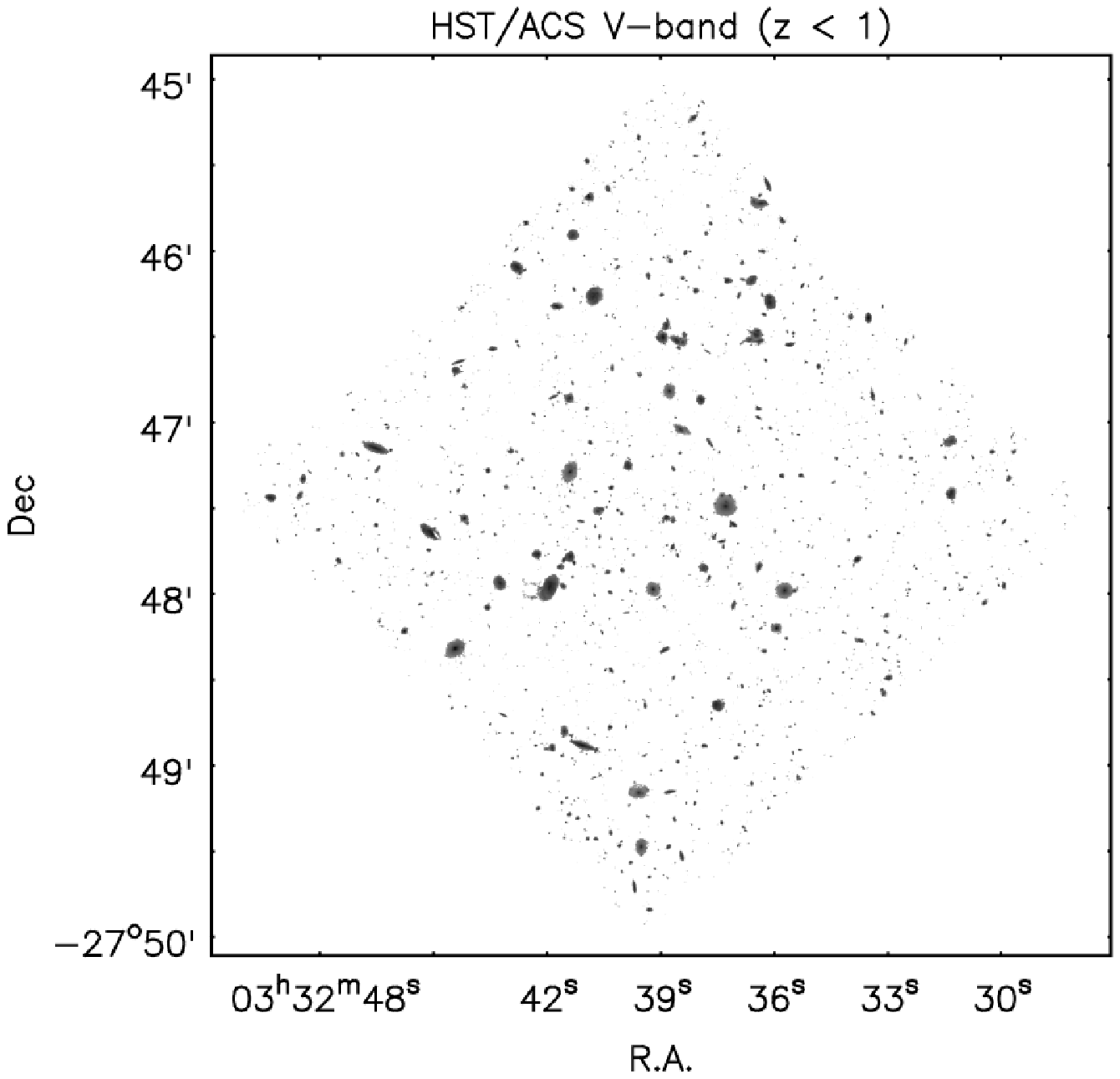}
\end{minipage}
\begin{minipage}{0.5\linewidth}
\centering
\includegraphics[trim = 0.25cm 0cm 0.1cm  0.1cm, clip,width=\textwidth]{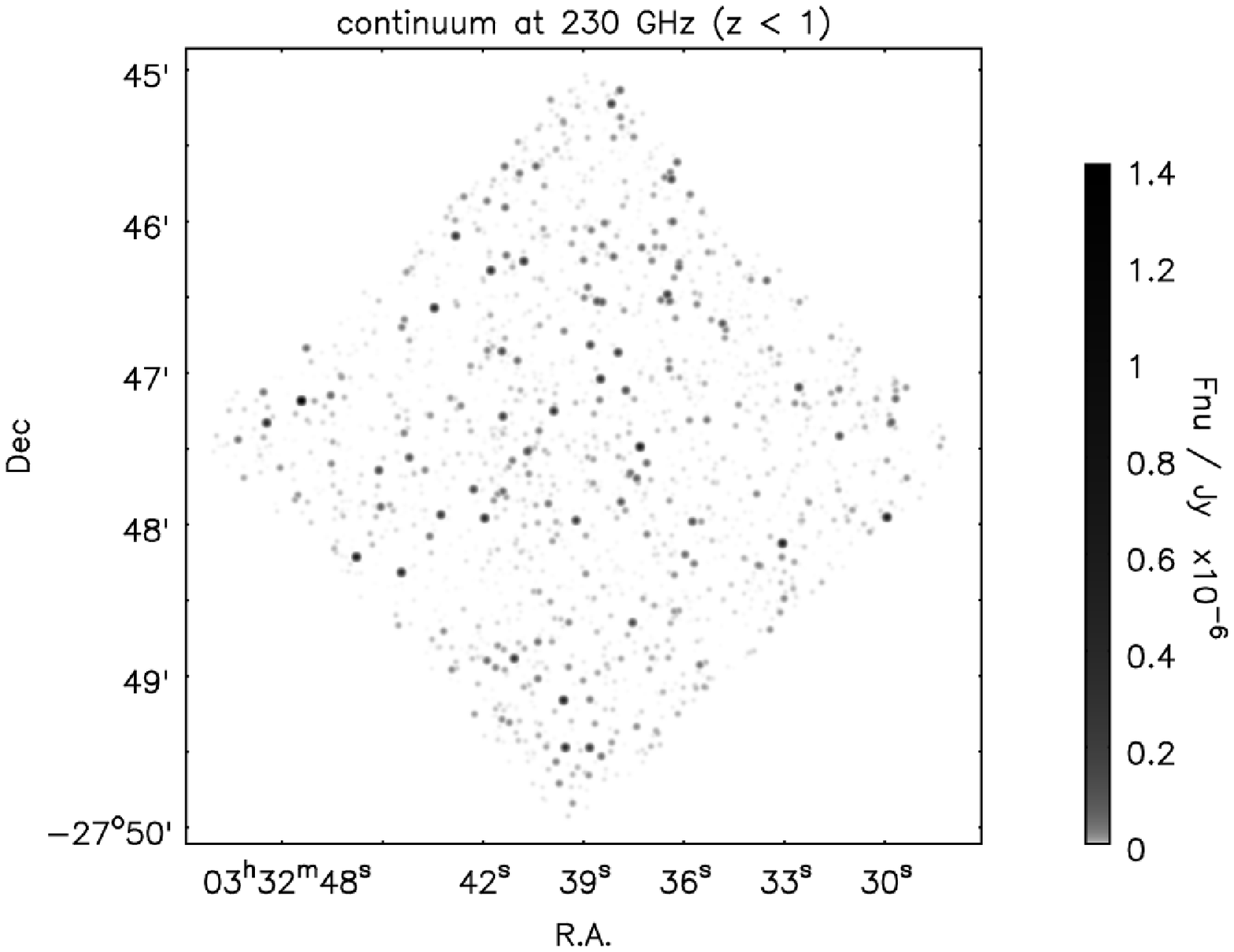}
\end{minipage}
\begin{minipage}{0.5\linewidth}
\centering
\includegraphics[trim = 0.25cm 0cm 0.05cm  0.1cm, clip,width=0.79\textwidth]{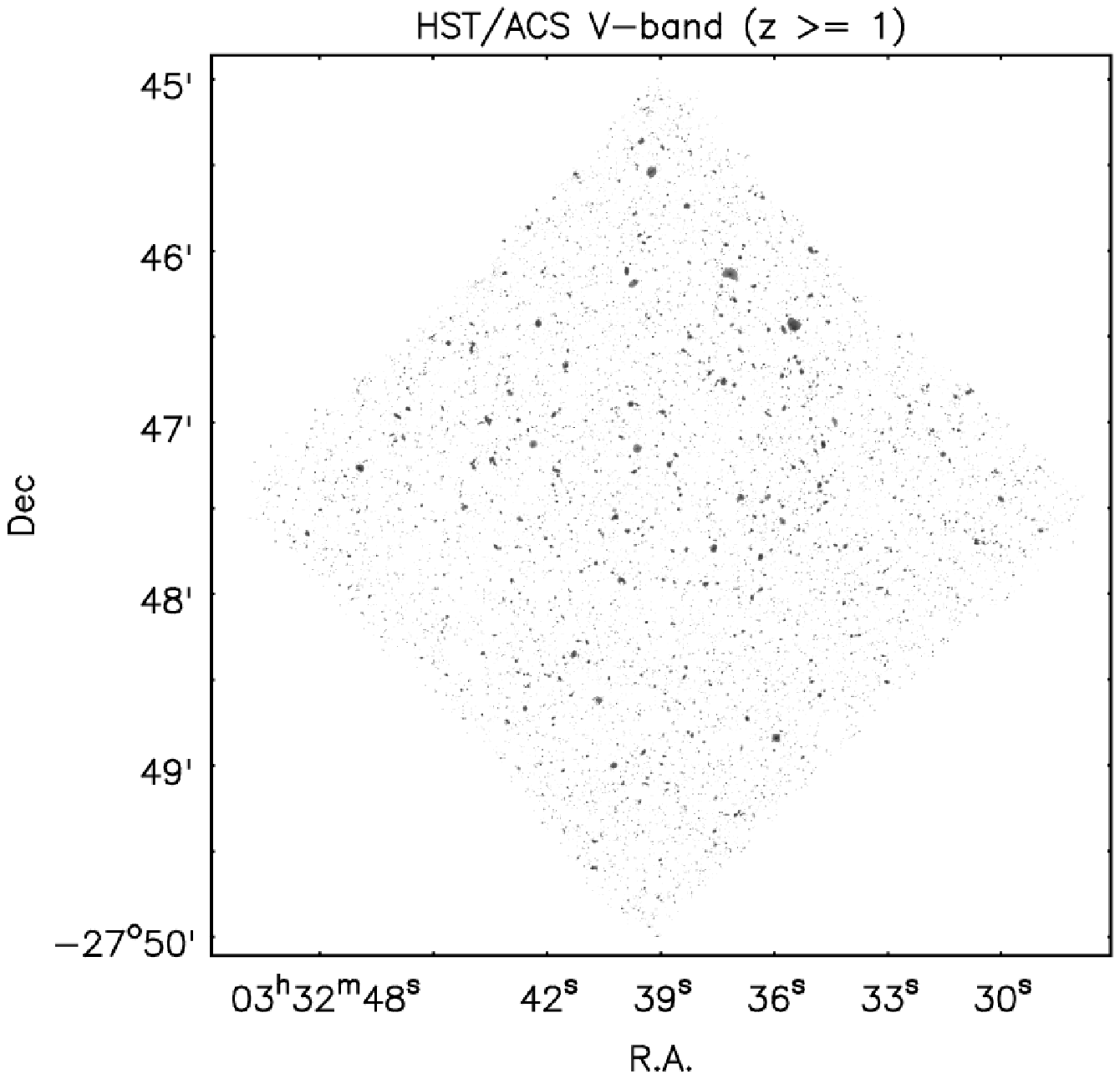}
\end{minipage}
\begin{minipage}{0.5\linewidth}
\centering
\includegraphics[trim = 0.25cm 0cm 0.1cm  0.1cm, clip,width=\textwidth]{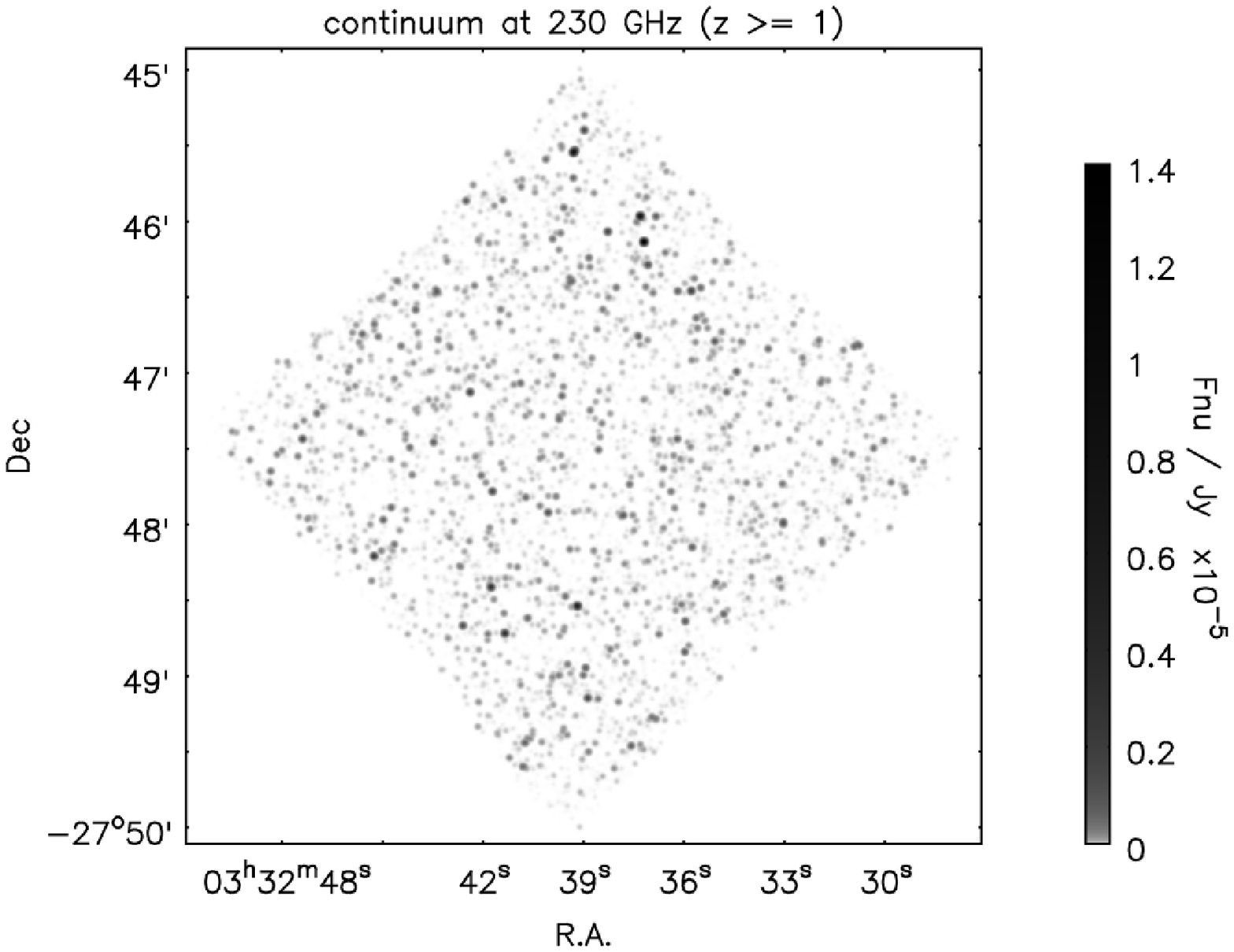}
\end{minipage}
\caption{Comparison between the UDF observed in the $V$-band (left-hand panels) and at 230~GHz (right-hand panels; using our predictions); top panels: whole sample; middle panels: galaxies with $z<1$; bottom panels: galaxies with $z\ge1$.
The $V$-band image was generated using ACS F606W image of the UDF (from the HST archive); the 230-GHz image was generated assuming point sources and convolving with the ALMA synthetic beam in the compact configuration, $1.5$~arcsec. The greyscale bar shows the predicted (sub-)mm fluxes of the galaxies; no noise is included in either the optical or (sub-)mm maps. We note that these maps do not include the SMG galaxy detected in the UDF as part of the LESS survey (LESSJ0333243.6-274644).}
\vspace{0.5cm}
\label{udf_maps}
\end{figure*}

As an example, in Fig.~\ref{udf_maps}, we plot continuum map of the UDF at 230~GHz (ALMA band 6, PdBI/NOEMA band 3) using our predictions (right-hand panels). Such images can be directly compared to future deep fields performed with ALMA or other facilities. We note that this figure is based solely on our optically-based predictions, and so they are missing the only known bright (sub-)mm source in the UDF:  LESSJ0333243.6-274644.
The top panels of Fig.~\ref{udf_maps} show our full sample, and we then divide the sample in two redshift bins
($z<1$; middle panels) and ($z\ge1$; bottom panels). This illustrates the differences between galaxy detections as a function of redshift between
the optical and the (sub-)mm, in particular that we expect galaxies to be relatively brighter in the (sub-)mm at
high redshift thanks to the negative k-correction. Therefore, we will be able to detect `normal' galaxies out to higher redshifts in the (sub-)mm with
new facilities, as we discuss in more detail in Section~\ref{sect:discussion}.

\begin{figure*}
\begin{center}
\includegraphics[width=0.9\textwidth]{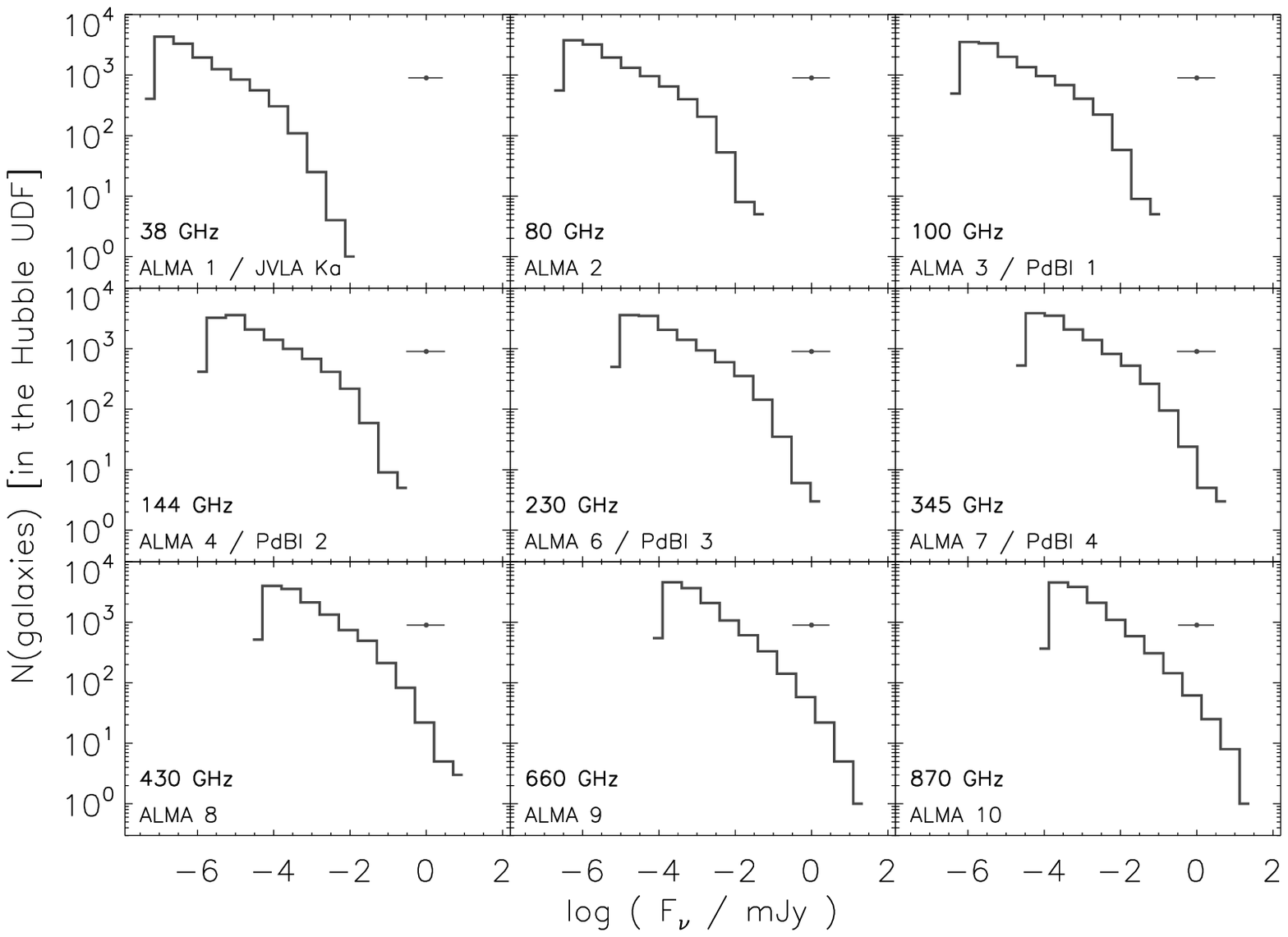}
\caption{Distribution of the predicted continuum flux densities per flux density bin in different (sub-)millimeter bands for all the galaxies in our sample. The frequency is indicated in the bottom-left corner of each panel, as well as the corresponding bands in different observatories. The median confidence range for the continuum flux in each band is plotted in the upper right-hand corner of each panel.}
\label{cont_fluxes_hist}
\end{center}
\end{figure*}

In Fig.~\ref{cont_fluxes_hist}, we plot the distribution of the predicted continuum flux densities of all the galaxies in our sample in all current and future ALMA, JVLA and IRAM PdBI/NOEMA bands from 38~GHz to 870~GHz. We plot the expected number of galaxies per flux bin in the total UDF area in each band. The distribution of fluxes peaks at higher fluxes in the highest frequency ALMA bands, because, even taking into account $k$-correction effects, these bands sample the emission from the galaxies closer to the peak of the dust SED. In Section~\ref{sect:discussion}, we discuss the feasibility of performing a blind survey of the UDF with ALMA at full operation, and use these predicted fluxes, combined with the projected ALMA sensitivities, to obtain an estimate of the expected number of continuum detections.

\subsection{CO and \cii\ line predictions}
\label{sect:lines}

ALMA and JVLA will detect CO and \cii\ line emission from high-redshift galaxies, which will allow us to determine redshifts, molecular gas reservoirs, and dynamical masses (e.g.~\citealt{Solomon2005,Daddi2010b,Genzel2010,Tacconi2010,Walter2011}). The rest-frame frequency, $\nu_\mathrm{rest}$, of the CO lines corresponding to $J \rightarrow J-1$ transitions from $J=1$ to $J=7$ and of the \cii\ line are given in Table~\ref{lines_z}. The observed frequency of each line varies with redshift as $\nu_\mathrm{obs}=\nu_\mathrm{rest}(1+z)^{-1}$. In Fig.~\ref{co_detect} (Appendix~\ref{appendix_co_lines}), we plot the observed frequency of the seven first CO transitions, CO(1--0) (i.e. $J=1$) to CO(7--6) (i.e. $J=7$) and of the \cii\ line as a function of redshift, with the frequency ranges covered by each ALMA, PdBI/NOEMA and JVLA band shaded in grey and green. In Table~\ref{lines_z} (Appendix~\ref{appendix_co_lines}), we explicitly list the redshift ranges where the CO lines and \cii\ are observable in each band. It is clear that the lowest frequency bands, such as JVLA K, Ka and Q will be crucial to probe the low $J$ CO transitions, in particular the CO(1--0) line, in $z>1$ galaxies. All the other PdBI/NOEMA and ALMA bands will potentially detect higher $J$ CO transitions at different redshifts, depending on the excitation state of the gas in galaxies. The highest-frequency ALMA bands will not only sample high-$J$ CO lines at low redshifts and the continuum dust emission nearest to its peak, as mentioned in the previous section, but also the \cii\ line out to high redshifts.

In this section, we attempt to predict the CO and \cii\ line fluxes for the galaxies in our UDF sample using empirical relations that relate line luminosities with the infrared luminosity of the galaxies, for which we have a statistical estimate from our SED fits (Section~\ref{sect:sed_results}).

\subsubsection{CO emission}
\label{sect:co}

\begin{figure*}
\begin{center}
\includegraphics[width=0.75\textwidth]{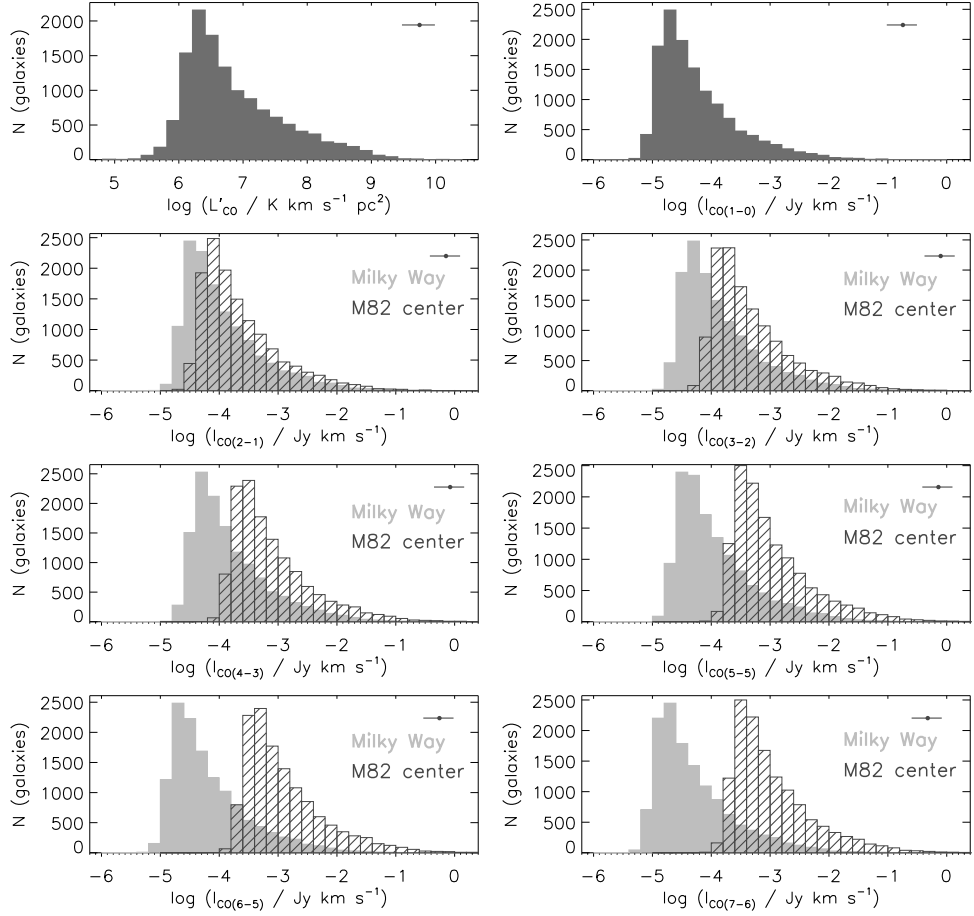}
\caption{Top panels: distribution of total CO line luminosity $L^{\prime}_\mathrm{CO}$ (top left) and velocity-integrated flux of the CO(1--0) line (top right) for all our sample. Other panels (starting at second row): distribution of the predicted velocity-integrated CO line fluxes for transitions $J=$ 2, 3, 4, 5, 6, and 7. These fluxes are computed from the predicted CO(1--0) line, using the CO SLEDs of \protect\cite{Weiss2007} for the Milky Way (light gray histogram) and the center of M~82 (dark gray, dashed histogram). These two extreme CO SLEDs should bracket a large realistic range of possible CO excitations of star-forming galaxies (e.g.~\citealt{Weiss2007}).}
\label{coflux}
\end{center}
\end{figure*}

The CO line luminosity of galaxies depends on various factors, such as the gas heating by starbursts, AGN, and the cosmic microwave background at high redshifts (e.g.~\citealt{Combes1999,Obreschkow2009}, da Cunha et al., in prep.), as well as the clumpiness and metallicity of the gas (e.g.~\citealt{Obreschkow2009}).
In this section, for simplicity, we predict the CO line luminosity of the galaxies in our sample using simple, empirically calibrated prescriptions.
It has been found for a wide range of galaxy types, both in the local and high-redshift Universe, that the CO line luminosity of star-forming galaxies correlates with their infrared luminosity (e.g.~\citealt{Solomon2005}, \citealt{Genzel2010}, \citealt{Daddi2010}).

The following relation between CO line luminosity and far-infrared luminosity was derived by \cite{Daddi2010} for BzK galaxies, i.e. gas-rich star-forming disks at high redshifts:
\begin{equation}
\log(\lir) = 1.13 \log(\lpco) + 0.53 \,,
\label{lir_lco}
\end{equation}
where \lir\ is the infrared luminosity (integrated between 8 and 1000~\mic) in \lsun\ and \lpco\ is the CO(1--0) line luminosity in K~km~s$^{-1}$~pc$^{2}$.
We obtain an estimate of \lpco\ using this equation and the statistical estimate on \lir\ obtained for the SED fits of our galaxies; the resulting distribution of \lpco\ for the whole sample is plotted in the top left-hand panel of Fig.~\ref{coflux}. We chose this empirical calibration between \lir\ and \lpco\ because our physical parameter estimates in Section~\ref{sect:properties} indicate that most of these galaxies would be comparable to normal, `main-sequence' star-forming disks, with typical infrared luminosities $\lir \lesssim 10^{11}~\lsun$. We note that eq.~\ref{lir_lco} is similar to the relation found by \cite{Genzel2010} for isolated, star-forming galaxies out to $z=2$. Other calibrations of this relation have been derived which include more extreme galaxies such as starbursts, mergers, and AGN (e.g.~\citealt{Solomon2005,Riechers2006}). When including these extreme galaxies, the relation between infrared and CO line luminosity becomes steeper, e.g. \cite{Solomon2005} find $\log(\lfir)=1.7\log(\lpco)-5.0$. Using this steeper relation for the infrared luminosity range of our galaxies ($\ldust \lesssim 10^{11}$~\lsun) would result in CO line luminosities over one order of magnitude higher than those predicted using Eq.~\ref{lir_lco} for the \lir\ range of the galaxies in our sample. We discuss the implications of using these different assumptions for the predicted number of CO line detections in Section~\ref{sect:discussion}. 

From \lpco\ (computed using Eq.~\ref{lir_lco}), we then compute the corresponding flux of the CO(1--0) line, $S_\nu^\mathrm{CO(1-0)}$, using (e.g.~\citealt{Solomon2005}):
\begin{equation}
\lpco = 3.25\times10^7 S_\nu^\mathrm{CO(1-0)} \, \Delta v \, \nu^{-2}_\mathrm{obs} \, D_L^2 (1+z)^{-3} \,,
\label{sco}
\end{equation}
where $S_\nu^\mathrm{CO(1-0)}$ is the flux density in Jy, $\Delta v$ is the line width in km~s$^{-1}$ (the velocity-integrated flux of the line is $I_\mathrm{CO(1-0)}=S_\nu^\mathrm{CO(1-0)} \, \Delta v$), $\nu_\mathrm{obs}$ is the observed frequency of the line in GHz, and $D_L$ is the luminosity distance in Mpc. We assume a typical line width of 300~km~s$^{-1}$, consistent with typical line-widths measured in high-redshift star-forming galaxies (e.g.~\citealt{Daddi2010,Genzel2010,Tacconi2010}). In the top right-hand panel of Fig.~\ref{coflux}, we plot the distribution of the velocity-integrated flux of the CO(1--0) line for all the galaxies in our sample computed using equation~\ref{sco}.

The fluxes of higher transition CO lines depend highly on the excitation of the CO gas in galaxies. Different physical conditions in the gas produce different CO spectral line energy distributions (SLEDs; e.g.~\citealt{Weiss2007}), which translate into different ratios between the CO(1--0) line and the higher $J$ lines. To compute the predicted flux of the CO(2--1), CO(3--2), CO(4--3), CO(5--4), CO(6--5) and CO(7--6) lines, we assume two possible CO SLEDs from \cite{Weiss2007}: the Milky Way CO SLED and the M~82 center CO SLED. These two cases correspond to very low and high excitation of the gas, respectively, and should bracket a large realistic range of possible physical conditions in star-forming galaxies. In the six bottom panels of Fig.~\ref{coflux}, we show the distribution of expected velocity-integrated CO line fluxes for the galaxies in our sample, and compare the predictions for these two excitation scenarios. For each CO line, these two extreme excitations should bracket the full range of line fluxes expected for our sample of star-forming galaxies (as supported by observations of multiple CO lines in a wide range of systems from local quiescent galaxies to high-redshift QSOs; see e.g. \citealt{Weiss2007}). The difference between the predictions of CO line fluxes using these two CO SLEDs increases with increasing $J$: the higher-$J$ CO lines are stronger in the case of the M~82 center SLED, which corresponds to a higher CO excitation. In Table~\ref{line_predictions_table}, we provide the predicted CO line fluxes for all the galaxies in our sample (including both CO excitation scenarios).

\begin{figure*}
\begin{minipage}[t]{\linewidth}
\centering
\includegraphics[width=0.7\textwidth]{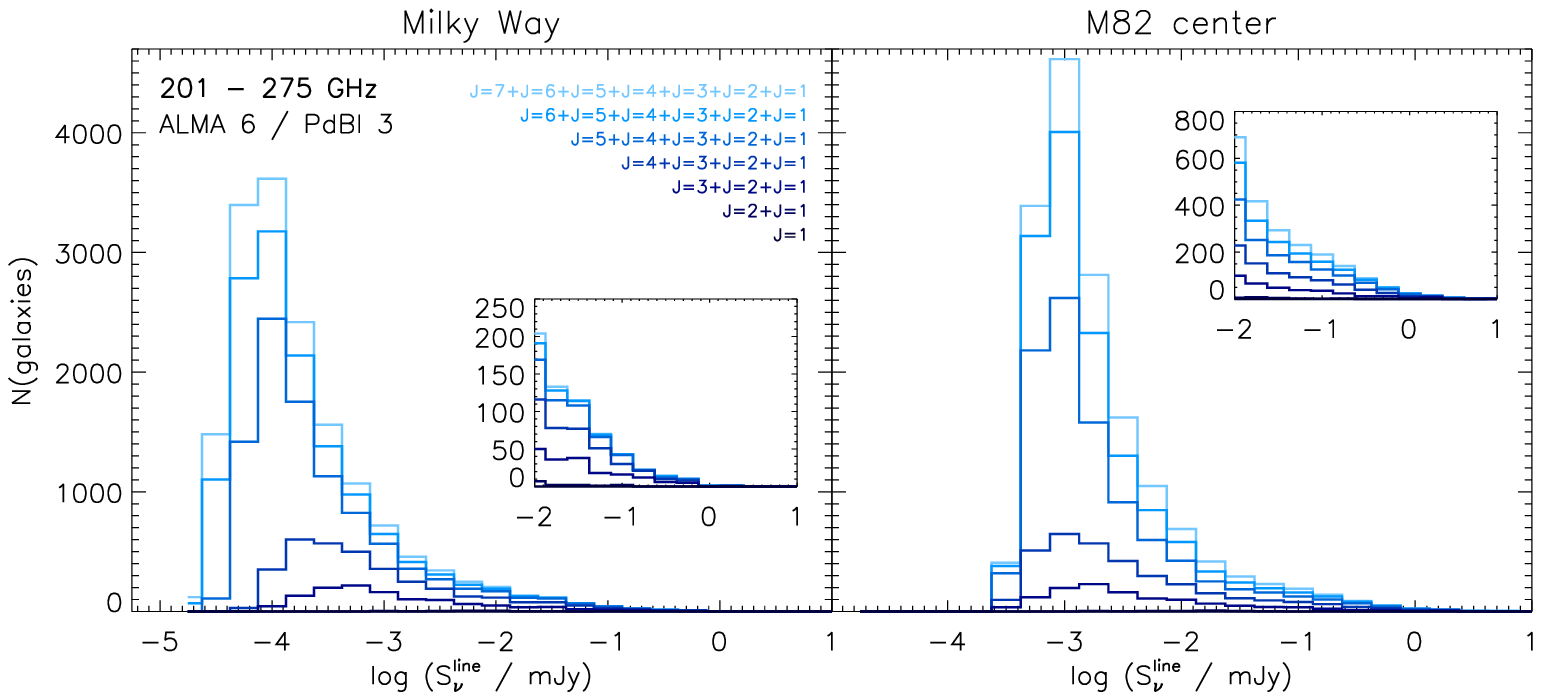}
\end{minipage}
\begin{minipage}[t]{\linewidth}
\centering
\includegraphics[width=0.7\textwidth]{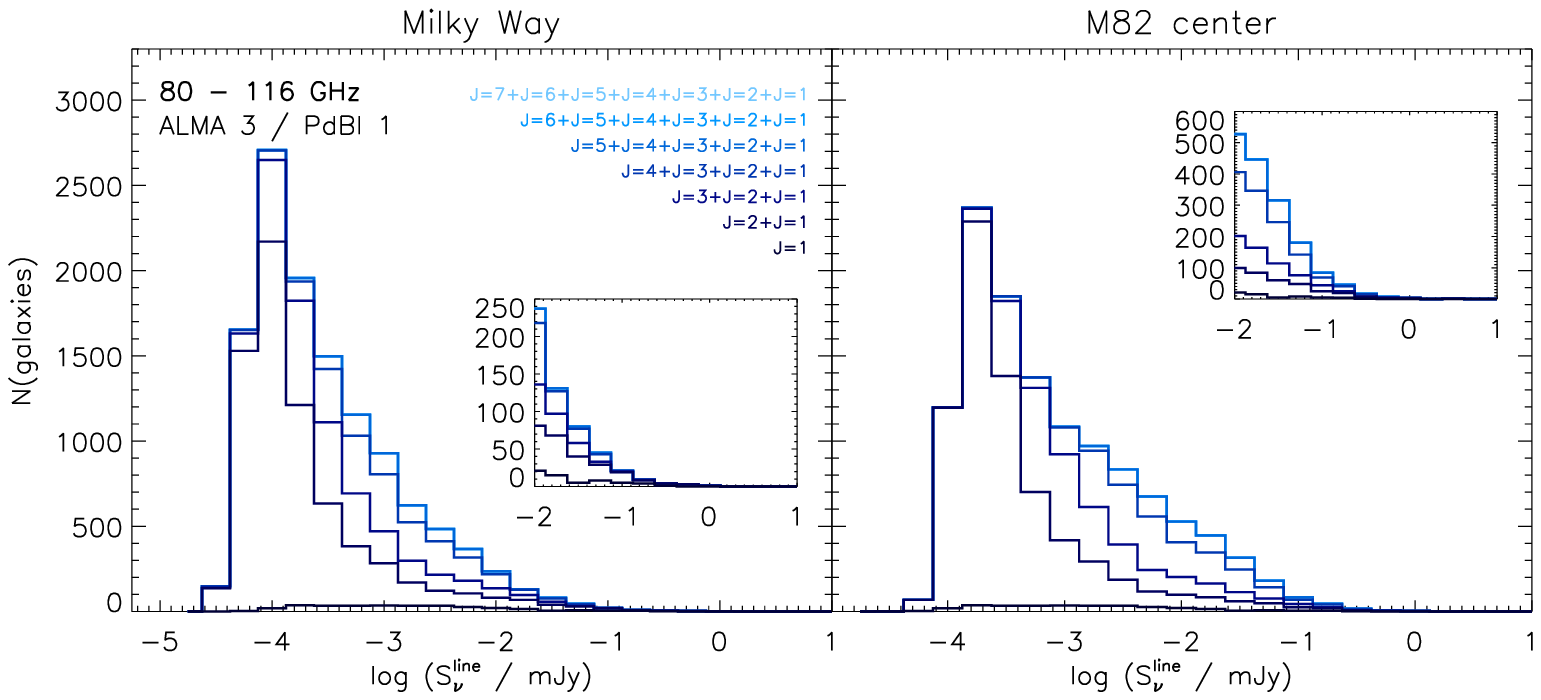}
\end{minipage}
\begin{minipage}[t]{\linewidth}
\centering
\includegraphics[width=0.7\textwidth]{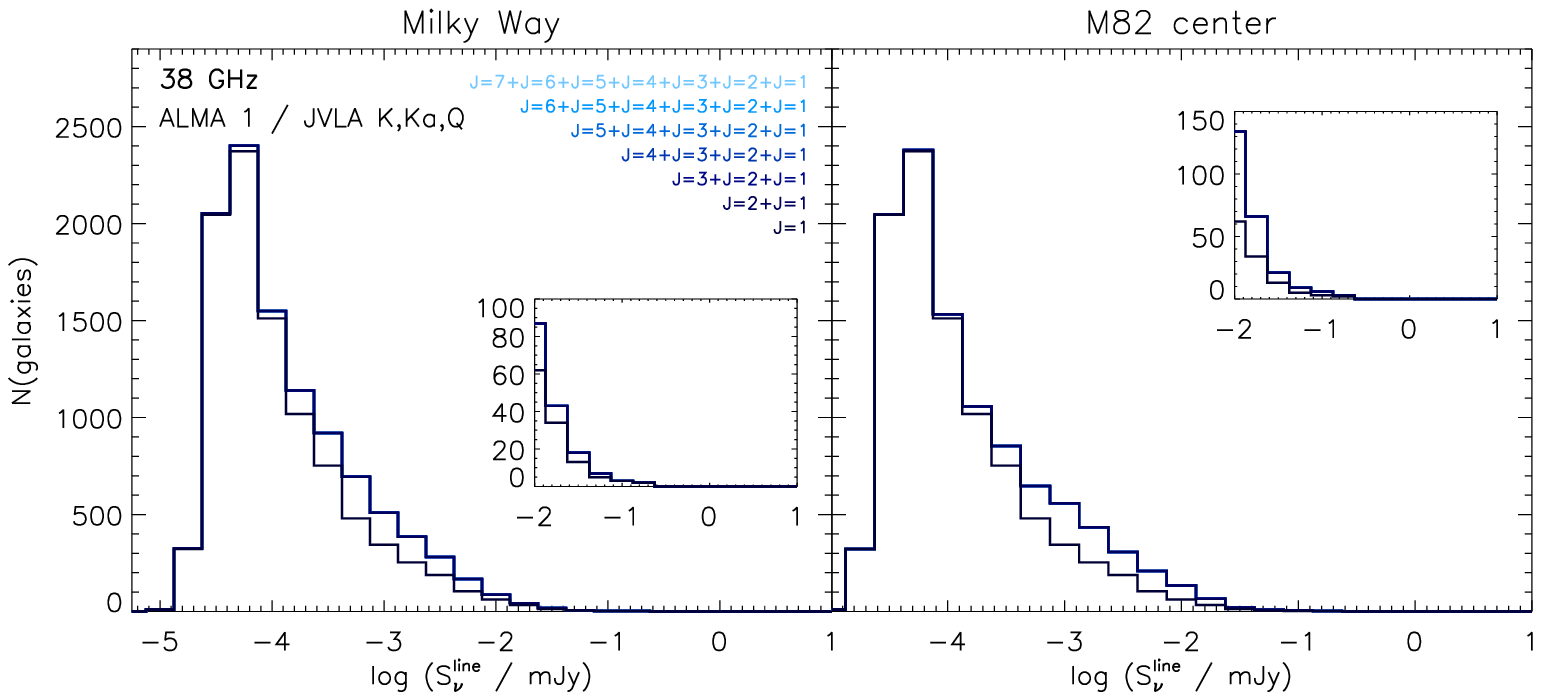}
\end{minipage}
\caption{Number of expected CO line detections per line flux bin for three frequency ranges and two different molecular gas excitation scenarios. The line fluxes are computed assuming a line width of 300~km~s$^{-1}$. {\em Top:} frequency between 201 and 275~GHz (ALMA band 6 and PdBI/NOEMA band 3); {\em Middle:} frequency between 80 and 116~GHz (ALMA band 3 and PdBI/NOEMA band 1); {\em Bottom:} frequency between 18 and 50~GHz (ALMA band 1 and JVLA bands K, Ka and Q).
The left-hand and right-hand panels assume a Milky Way and M82 CO spectral line energy distribution, respectively. The lowest, darkest color histograms show the distribution of CO(1--0) fluxes, the next (lighter-colored) histogram shows the joint distribution of CO(1--0) and CO(2--1) fluxes, i.e. the distribution of CO(2--1) can be inferred from the increment in the histogram relatively to the histogram below, and so on until the lightest-colored histogram, which shows the sum of the distributions of all the CO line fluxes from $J=1$ to $J=7$.}
\vspace{0.25cm}
\label{co_detect_flux}
\end{figure*}

To predict the number of expected CO line detections given a certain flux limit in various (sub-)millimeter bands (ALMA, JVLA or PdBI/NOEMA), we can build the expected distribution of CO line fluxes in each band based on the predictions described above. First, for each frequency band listed in Table~\ref{lines_z}, we retain galaxies for which the redshift falls in a range where one of the CO lines can be observed (these ranges are listed in Table~\ref{lines_z}; see also Fig.~\ref{co_detect}). Then, we compute the expected line fluxes, $S_\nu^\mathrm{line}$ in Jy, for each galaxy, by assuming a typical line width of $\Delta v = 300$~km s$^{-1}$ (using eq.~\ref{sco} to the get flux of the CO(1--0) line and the CO SLEDs to get the higher-$J$ lines). In Fig.~\ref{co_detect_flux}, as examples, we plot the distribution of line fluxes from CO(1--0) to CO(7--6) if one were to fully cover the bands from 201 to 275~GHz (ALMA band 6, PdBI/NOEMA band 3), 80 to 116~GHz (ALMA band 3, PdBI/NOEMA band 1) and 18 to 50~GHz (ALMA band 1 / JVLA bands K, Ka and Q). We show the CO line fluxes corresponding to two gas excitation scenarios (i.e.~CO SLEDs): Milky Way-type (left-hand panels) and M82-type (right-hand panels). For example, in the frequency range 80 to 116~GHz, we plot the distribution of CO(1--0) line fluxes only of galaxies with redshifts $z\leq0.44$, for which the CO(1--0) line would be redshifted to that frequency band (Table~\ref{lines_z}). Similarly, for the distribution of CO(2--1) line fluxes in that band, we include only galaxies with $0.99 \leq z \leq 1.88$, and so on until the CO(7--6) line. For clarity, the distributions plotted in Fig.~\ref{co_detect_flux} are cumulative: the darkest-colored histogram shows the number of galaxies per line flux bin for the CO(1--0) line, the next, lighter histogram shows the number of galaxies per line flux bin for the CO(1--0) and CO(2--1) line, the next histogram adds the number of galaxies per line flux bin for the CO(3--2) line, etc. That is, the lightest-colored histogram shows the total number of galaxies per line flux bin when including all CO lines from $J=1$ to $J=7$. In general, Fig.~\ref{co_detect_flux} shows that the number of galaxies in the highest flux bins is largest if the CO SLED is M82-like, as expected, so the number of detections given a certain flux limit greatly depends on the CO excitation (as discussed in Section~\ref{discussion_line}).

\subsubsection{[CII] emission}
\label{sect:cii}

The \cii\ line at 158~\mic\ is the main cooling line of the ISM in galaxies, and it arises mainly from photo-dissociation regions -- at the interface between the ionized gas and the neutral and molecular gas -- which are typically associated with star-forming regions. This line is therefore one of the main far-infrared tracers of star formation in galaxies \citep{Stacey1991,Boselli2002,Stacey2010,deLooze2011}. Since this is the brightest far-infrared line, it will be readily detected in deep observations with ALMA, particularly using the highest-frequency bands.

\begin{figure*}
\begin{center}
\includegraphics[width=0.85\textwidth]{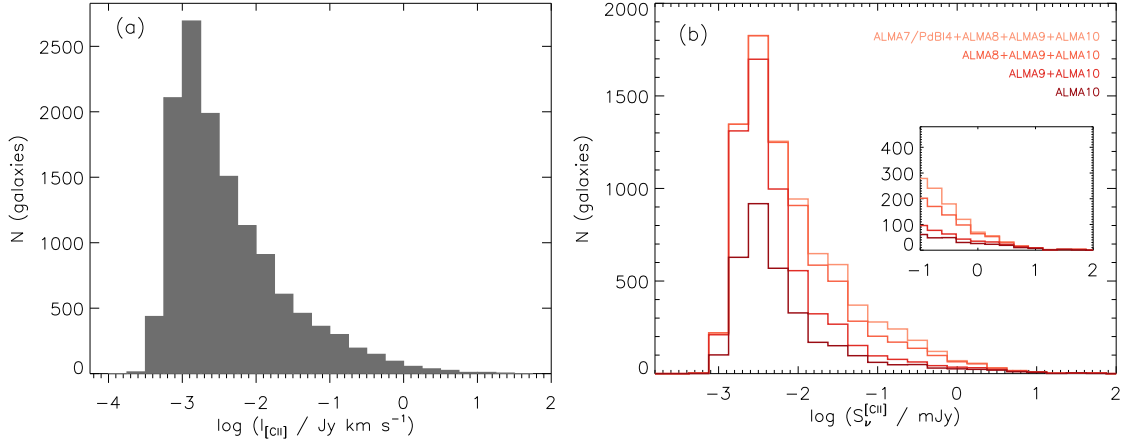}
\caption{(a) Distribution of the velocity-integrated flux of the $\cii$ line, $I_\mathrm{[CII]}$, for all the galaxies in our sample. (b) Distribution of the number of detections as a function of line flux, $S_\nu^\mathrm{[CII]}$ (computed assuming a 300~km~s$^{-1}$ line width), expected in the highest frequency bands available for ALMA and PdBI/NOEMA.}
\label{cii_detect}
\end{center}
\end{figure*}

We rely on previous observational studies of the ratio of \cii\ line to far-infrared luminosity ($L_\mathrm{[CII]} / \lfir$) of galaxies to estimate the \cii\ line luminosity for each galaxy in our sample.
The ratio $L_\mathrm{[CII]} / \lfir$ of normal star-forming galaxies varies typically between 1 and 0.1\%, and has been shown to anti-correlate with dust heating intensity (e.g.~\citealt{Brauher2008}). 
Here, for simplicity, and considering the relatively large error bars on our \lfir\ estimates, we adopt a constant ratio of $\log(L_\mathrm{[CII]} / \lfir)=-2.5$, which corresponds to the average value for normal star-forming galaxies with low far-infrared luminosities $\lir \lesssim 10^{11}~L_\odot$ (i.e. similar to the galaxies in our sample) and average dust heating \citep{Boselli2002,Brauher2008,Gracia2011,Cox2011}. In Fig.~\ref{cii_detect}(a), we plot the distribution of the expected velocity-integrated flux of \cii, $I_\mathrm{[CII]}=S_\nu^\mathrm{[CII]} \, \Delta v$ for all the galaxies in our sample, computed using eq.~\ref{sco}:

\begin{equation}
\lcii = 1.04\times10^{-3} S_\nu^\mathrm{[CII]} \, \Delta v \, \nu_\mathrm{rest} \, D_L^2 (1+z)^{-1} \,,
%\lpcii = 3.25\times10^7 S_\nu^\mathrm{[CII]} \, \Delta v \, \nu^{-2}_\mathrm{obs} \, D_L^2 (1+z)^{-3} \,,
\label{scii}
\end{equation}
where \lcii\ is the \cii\ line luminosity in \lsun, $S_\nu^\mathrm{[CII]}$ is the line flux in Jy, $\Delta v$ is the velocity dispersion in km~s$^{-1}$, $D_L$ is the luminosity distance, and $\nu_\mathrm{rest}=1900.54$~GHz. At $z<5$, the observed frequency of the \cii\ line falls in the highest frequency ALMA bands, namely bands 7, 8, 9 and 10 (Fig.~\ref{co_detect}). In Fig.~\ref{cii_detect}(b), we plot the distribution
of expected detections in these bands as a function of \cii\ line flux (computed assuming $\Delta v=300$~km~s$^{-1}$).
In Table~\ref{line_predictions_table}, we provide the predicted \cii\ line fluxes for all the galaxies in our sample.

%*****************************************************************************
\section{Possible deep field strategies with ALMA}
\label{sect:discussion}

Based on the results presented in the previous sections, we now discuss the feasibility of carrying out a deep field survey of the Hubble UDF with ALMA at Full Operations (i.e. using 50 antennas).
In the following, we will assume a total time of 500~hours for the full survey, and investigate the possible setups of such a survey to maximize the redshift coverage and number of galaxy detections.

One immediate drawback of ALMA as a survey instrument is that the primary beam size, which is driven by the size of the antennas, is relatively small in all bands: the primary beam size is given, to first order, by $20^{\prime\prime}.3\times300/(\nu/\mathrm{GHz})$, where $\nu$ is the observing frequency. Therefore, even a relatively small area field as the Hubble UDF ($3.45^\prime\times3.45^\prime$, i.e. a total area of 11.9~arcmin$^2$), will be hard to cover with ALMA, and will require significant mosaicking. It is beyond the scope of this paper to go into the technical details of how such a mosaic should be set up. In the following, for illustration purposes, we simply consider a total on-source time of 500 hours for the UDF, which we divide in different pointings to cover the area of the UDF -- we do not include overheads and mosaicking details (such as degrading sensitivities inside the beam etc.).

\subsection{Continuum detections}
\label{discussion_continuum}

\begin{deluxetable*}{c c c c c c c c c c }
%\rotate
\tablecolumns{10}
\tablecaption{Summary of a possible observational set-up to observed the Hubble UDF with ALMA in full operations with a total on-source time of 500 hours.}
\tablewidth{0pt}
\tablehead{
\colhead{} &
\colhead{} &
\colhead{} &
\colhead{} &
\colhead{number of} &
\colhead{time per} &
\colhead{time per} &
\colhead{} &
\colhead{} &
\colhead{number of} \\
\colhead{Band} &
\colhead{frequency} &
\colhead{primary beam} &
\colhead{number of} &
\colhead{frequency} &
\colhead{pointing\tablenotemark{b}} &
\colhead{frequency setting} &
\colhead{$\sigma_\mathrm{cont}$\tablenotemark{c}} &
\colhead{$\sigma_\mathrm{line}$\tablenotemark{d}} &
\colhead{$3\sigma$ continnuum} \\
\colhead{} &
\colhead{/ GHz} &
\colhead{/ arcsec} &
\colhead{pointings\tablenotemark{a}} &
\colhead{settings} &
\colhead{/ h} &
\colhead{/ h} &
\colhead{/ mJy} &
\colhead{/ mJy} &
\colhead{detections\tablenotemark{e}}
}
\startdata
ALMA 1 & 38 & 140 & 3 & 2 &167 & 83.5 & $3.7\times10^{-4}$ & $7.6\times10^{-3}$ & 36\\
ALMA 2 & 80 & 76 & 10 & 3& 50 & 16.7 & $9.9\times10^{-4}$ & $1.7\times10^{-2}$ &155\\
ALMA 3 & 100 & 62 & 15 & 4 & 33 & 8.25 & $1.5\times10^{-3}$ & $2.7\times10^{-2}$ & 221\\
ALMA 4 & 144 & 43 & 30 & 5 & 17 & 3.40 & $2.5\times10^{-3}$ & $4.2\times10^{-2}$ & 363\\
ALMA 6 & 230 & 26 & 81 & 8 & 6.2 & 0.78 & $5.1\times10^{-3}$ & $8.4\times10^{-2}$ & 601\\
ALMA 7 & 345 & 18 & 169 & 13 & 3.0 & 0.23 & $1.3\times10^{-2}$ & $2.3\times10^{-1}$ & 522\\
ALMA 8 & 430 & 14 & 278 & 15 & 1.8 & 0.12 & $8.0\times10^{-2}$ & $1.3\times10^{0}$ & 136\\
ALMA 9 & 660 & 9.3 & 630 & 14 & 0.8 & 0.06 & $2.4\times10^{-1}$& $3.1\times10^{0}$ & 84\\
ALMA 10 & 870 & 1.1 & 1080 & 21 & 0.5 & 0.03 & $7.1\times10^{-1}$&$9.5\times10^{0}$ & 40\\ 
\enddata
\label{continuum_detections}
\tablecomments{Sensitivities computed using the ALMA Sensitivity Calculator (50 antennas and default weather conditions).}
\tablenotetext{a}{Computed by dividing the total UDF area by the area of the primary beam in each band.}
\tablenotetext{b}{Computed by dividing 500 hours by number of pointings in each band.}
\tablenotetext{c}{8~GHz bandwidth.}
\tablenotetext{d}{300~km/s bandwidth.}
\tablenotetext{e}{Expected number of continuum detections over the whole UDF field for each band. The expected number of line detections is given in Table~\ref{line_detections}.}
\end{deluxetable*}

\begin{figure*}
\begin{center}
\includegraphics[width=0.985\textwidth]{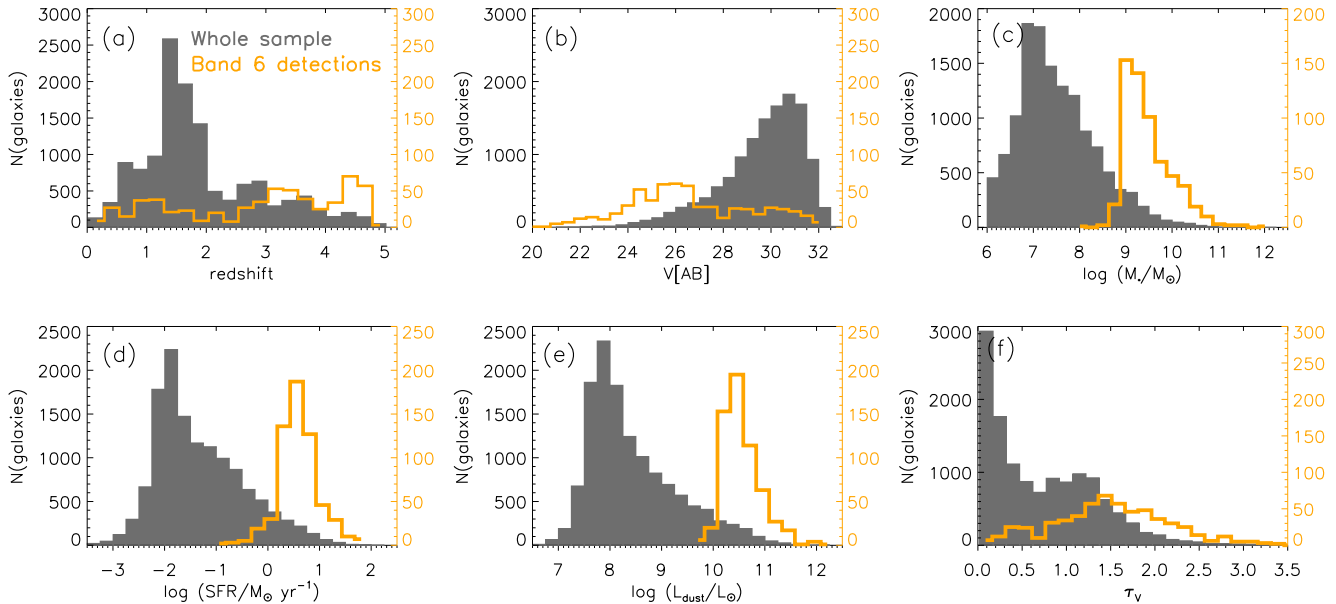}
\caption{Comparison between the distribution of redshift (a), $V$-band magnitude (b), stellar mass (c), star formation rate (d), dust luminosity (e) and $V$-band optical depth in stellar birth clouds (f) for our whole sample of galaxies (gray filled histograms),
and for the 601 galaxies expected to be detected at the 3$\sigma$-level in the continuum in ALMA band 6 at 230~GHz, following the observational setup described in Table~\ref{continuum_detections} (orange histograms, with numbers given on the right-hand y-axis). It is clear from these plots that the detected galaxies will be the most massive, star-forming and dust-rich galaxies in the UDF.}
\label{detections_band6}
\end{center}
\end{figure*}

For each ALMA band, given the area of the primary beam, we compute how many effective pointings are needed to cover the whole area of the UDF (Table~\ref{continuum_detections}). The number of pointings needed to cover the UDF increases from Band 1 to Band 10, from only 3 pointings to over 1000 pointings.
In this back-of-the-envelope calculation we take the integration time per pointing in each band as the total 500 hours divided by the required number of pointings, and then use the ALMA Sensitivity Calculator
(ASC)\footnote{http://almascience.eso.org/call-for-proposals/sensitivity-calculator} to compute the continuum sensitivity that can be reached in each integration time, assuming a 8~GHz bandwidth and the default weather conditions. In Table~\ref{continuum_detections}, we list the sensitivities and the expected number of 3$\sigma$ continuum detections in each band for the whole field, based on our continuum predictions of Section~\ref{sect:results_continuum}. The resulting number of predicted continuum detections is a combination of the intrinsic flux of each galaxy, the $k$-correction for each galaxy at each redshift, and the changing integration time per pointing due to varying beam size; for reference, 1 hour integration time corresponds to a rms of 4.81, 12.6 and 483~$\mu$Jy at 38~GHz (ALMA 1), 230~GHz (ALMA 6) and 870~GHz (ALMA 10), respectively.
Table~\ref{continuum_detections} shows that the number of expected continuum detections is maximum for band 6 at 230~GHz (601 detections), with a large number of predicted detections also in band 3 at 100~GHz (221) band 4 at 144~GHz (363) and band 7 at 345~GHz (522). In Figs.~\ref{redshift} and \ref{detections_band6}, we compare the distributions of the properties of the 601 galaxies that would be detected at the 3$\sigma$-level in the ALMA band 6 to those of the original sample. We note that the redshift distribution of these sources is relatively flat, thanks to the negative $k$-correction in the sub-mm. Also not surprisingly, these figures show that we expect to detect only the most massive, highly star-forming and dusty galaxies in our sample. These galaxies are still over an order of magnitude star-forming and dusty than classic SMGs detected in blind `pre-ALMA' (sub-)mm surveys, which have typically star formation rates $\gtrsim 100~\msun~$yr$^{-1}$ and dust luminosities $\gtrsim 10^{12}~\lsun$. The median detected galaxy has a stellar mass of $3\times10^9~M_\odot$, a star formation rate of $\sim5~M_\odot \mathrm{yr}^{-1}$, dust luminosity of $4\times10^{10}~L_\odot$, dust mass of $5\times10^7~M_\odot$ and $V$-band effective optical depth in stellar birth clouds of 1.6. This implies that the typical detected galaxy would be about 100 times more star-forming, more massive (in terms of stellar content) and more dusty, and about 10 times more obscured in the optical than the median of the whole UDF sample.

\subsection{Line detections}
\label{discussion_line}

\begin{table}\scriptsize
\begin{center}
\caption{Predicted number of 5$\sigma$ CO line detections in the Hubble UDF with ALMA, assuming a total on-source integration time of 500 hours,
for two different cases of CO line SLEDs: Milky Way and M~82 center (Section~\ref{sect:co}).}
\begin{tabular}{ l c c }
\tableline
\tableline
& {\bf ALMA 1 [31.3 -- 45~GHz]} & \\
& $5\sigma_\mathrm{line}=3.82\times10^{-2}$~mJy  & \\
\tableline
Line & Milky Way & M~82 center\\
\hline
CO(1--0) & 17 & 17 \\
CO(2--1) & 5 & 16 \\
\tableline
\tableline
& {\bf ALMA 2 [67 -- 90~GHz]} & \\
& $5\sigma_\mathrm{line}=8.55\times10^{-2}$~mJy & \\
\tableline
Line & Milky Way & M~82 center\\
\tableline
CO(1--0) & 21 & 21 \\
CO(2--1) & 7 & 16 \\
CO(3--2) & 1 & 38 \\
CO(4--3) & 2 & 14 \\
\tableline
\tableline
& {\bf ALMA 3 [84 -- 116~GHz] }& \\
& $5\sigma_\mathrm{line}=1.34\times10^{-1}$~mJy  & \\
\tableline
Line & Milky Way & M~82 center\\
\tableline
CO(1--0) & 5 & 5 \\
CO(2--1) & 15 & 37 \\
CO(3--2) & 0 & 17 \\
CO(4--3) & 0 & 37 \\
CO(5--4) & 0 & 11 \\
\tableline
\tableline
& {\bf ALMA 4 [125 -- 163~GHz] }& \\
& $5\sigma_\mathrm{line}=2.10\times10^{-1}$~mJy & \\
\tableline
Line & Milky Way & M~82 center\\
\tableline
CO(2--1) & 14 & 26 \\
CO(3--2) & 4 & 27 \\
CO(4--3) & 0 & 12 \\
CO(5--4) & 0 & 32 \\
CO(6--5) & 0 & 27 \\
CO(7--6) & 0 & 8 \\
\tableline
\tableline
& {\bf ALMA 6 [211 -- 275~GHz] }& \\
& $5\sigma_\mathrm{line}=4.22\times10^{-1}$~mJy & \\
\hline
Line & Milky Way & M~82 center\\
\tableline
CO(2--1) & 0 & 0 \\
CO(3--2) & 10 & 30 \\
CO(4--3) & 6 & 39 \\
CO(5--4) & 3 & 43 \\
CO(6--5) & 0 & 18 \\
CO(7--6) & 0 & 7 \\
\tableline
\tableline
& {\bf ALMA 7 [275 -- 373~GHz] }& \\
& $5\sigma_\mathrm{line}=1.15\times10^{0}$~mJy & \\
\hline
Line & Milky Way & M~82 center\\
\hline
CO(3--2) & 0 & 0 \\
CO(4--3) & 0 & 22 \\
CO(5--4) & 0 & 30 \\
CO(6--5) & 0 & 22 \\
CO(7--6) & 0 & 10 \\
\tableline
\label{line_detections}
\end{tabular}
\end{center}
\end{table}

\begin{table}
\begin{center}
\caption{Predicted number of 5$\sigma$ \cii\ line detections in the Hubble UDF with ALMA, assuming a total on-source integration time of 500 hours.}
\begin{tabular}{ l c c c }
\tableline
\tableline
 Band & Frequency range & 5$\sigma_\mathrm{line}$ & number of 5$\sigma$\\
 & / GHz & / mJy & \cii\ detections \\
\tableline
ALMA 7 & 275 -- 373 & 1.15 & 10\\
ALMA 8 & 385 -- 500 & 6.65 & 0\\
ALMA 9 & 602 -- 710 & 15.6 & 3\\
ALMA 10 & 787 -- 950 & 47.3 & 2\\
\tableline
\label{line_detections_cii}
\end{tabular}
\end{center}
\end{table}

We now discuss the possibility of detecting CO and \cii\ lines in the UDF using frequency scans in all ALMA bands. We compute the expected line sensitivities for a total on-source time of 500 hours on the UDF as in Section~\ref{discussion_continuum}, but taking into account the time needed to scan in band in frequency space. The frequency interval covered in each frequency setting is $\Delta\nu=8$~GHz. Therefore, the total number of settings required to scan a given ALMA band is the total frequency range of the band divided by $\Delta\nu$ (we note that the exact setup will depend on the sideband-separations in the various bands). The total time spent in each frequency setting is then the integration time per pointing divided by the required number of settings in each band. We assume the same mosaicking scheme of the UDF and use the same integration time per pointing in each band as for the continuum (Section~\ref{discussion_continuum}). The total number of frequency settings, time per setting and resulting line sensitivity $\sigma_\mathrm{line}$ (computed using the ASC and assuming a bandwidth of 300~km~s$^{-1}$ in order to resolve the lines in velocity space) are listed in Table~\ref{continuum_detections}. In Table~\ref{line_detections}, we show the predicted number of $5\sigma$ line detections in each band given the line sensitivities $\sigma_\mathrm{line}$ of Table~\ref{continuum_detections}, assuming both the Milky Way and the M\,82 CO SLED, as in Section~\ref{sect:co}. It is clear that with the integration times per frequency setting of Table~\ref{continuum_detections} we predict a relatively low number of line detections, compared with the predicted number of continuum detections. In bands 2 to 6, we predict a minimum of about 20 CO line detections in each band, when assuming the Milky Way CO SLED; around 100 detections are predicted if the CO is more excited. To get more line detections, one would need to go deeper, which likely implies a compromise with the area of the survey. For example, to reach a rms of 10~$\mu$Jy in band 6 (approximately 8 times deeper than reached in the ``default" survey set-up considered so far), which would yield at least between 200 and 800 detections over the whole UDF, one would need over 400~hours of effective integration time for each of the 81 pointings.

As mentioned in Section~\ref{sect:co}, if we use the \cite{Solomon2005} calibration to convert the infrared luminosities of our galaxies into CO line luminosities, we would obtain intrinsically brighter CO lines by about one order of magnitude. This would make the number of predicted CO line detections in the observational setup discussed here much higher than the numbers listed in Table~\ref{line_detections}. In the bands with most predicted detections, ALMA 2 to ALMA 6, the minimum number of detections (when assuming the Milky Way SLED) would increase from on average 22 to 116 in each band, and the maximum number of detections (when assuming the M\,82 SLED) would increase from on average 50 to 242 in each band.

We use the same method to estimate the number of expected \cii\ line detections in the highest frequency ALMA bands, based on our estimates of \cii\ line fluxes from Section~\ref{sect:cii}. We predict a total of 15 \cii\ 5$\sigma$-detections using the integration times per pointing and per frequency setting in each band listed in Table~\ref{continuum_detections} (Table~\ref{line_detections_cii}). In bands 8, 9 and 10, these numbers represent a low fraction ($\lesssim 0.1$ per cent) of the total number of galaxies in the observable redshift range because the large number of required pointings to cover the whole UDF implies a very short integration time per pointing and hence a relatively high rms. The minimum dust luminosities and star formation rates of the galaxies that can be detected in \cii\ using this observational setup are $5\times10^{11}$~\lsun\ and 10~\msun\ yr$^{-1}$, at the very high end of the distribution for the whole sample (e.g.~Fig~\ref{detections_band6}). With band 7, the number of detections is higher and because the rms is smaller thanks to a higher integration time, allowing us to go deeper and detect \cii\ emission from galaxies with dust luminosities as low as $1.4\times10^{11}$~\lsun.

%*****************************************************************************
\section{Discussion \& Conclusions}
\label{conclusion}

In this paper, we have presented empirical predictions of (sub-)millimeter continuum and CO/\cii\ line fluxes for a sample of 13,099 galaxies with redshifts up to $z=5$, selected in the deepest optical/near-infrared catalog of the Hubble Ultra Deep Field over 12~arcmin$^2$.
We have performed a self-consistent modelling of the observed optical/near-infrared spectral energy distributions of the galaxies, which relies on an energy-balance technique, and have allowed us to obtain Bayesian estimates of the total infrared luminosity and (sub-)millimeter continuum flux densities in several ALMA, JVLA and PdBI/NOEMA bands. We then combined the constraints on total infrared luminosity of our galaxies with empirical correlations to obtain estimates of their CO and \cii\ line luminosities. One advantage of our method is that it allows us to derive reliable confidence ranges for the physical parameters of the galaxies derived from SED fitting. Using fits to the complete ultraviolet to far-infrared SEDs of a sub-sample of galaxies in the UDF, we show that, even though our confidence ranges for the infrared luminosities and (sub-)millimeter continuum fluxes can be as large as one order of magnitude, our estimates are reliable, and the large confidence ranges reflect the inherent uncertainty in deriving infrared and (sub-)millimeter properties of galaxies from optical observations.

Our predictions rely mainly on two assumptions. First, that we can reliably predict the total infrared luminosity from the attenuated stellar emission, i.e. that our dust attenuation prescription is correct. We expect this to be true for moderately star-forming, optically-selected galaxies with moderate dust optical depths and infrared luminosities \citep{Charlot2000,daCunha2008,Daddi2005,Daddi2007,Reddy2006}. Another argument in favor of the relatively low typical optical depths of our galaxies is that, if we assumed infrared-to-optical ratios typical of very dust-obscured ULIRGs, we would strongly overestimate the extragalactic background light at 870\mic\ (Appendix~\ref{appendix_seds}). The second assumption is that the empirical relations between infrared luminosities and line emission, calibrated for low-luminosity galaxies only in the local Universe for obvious observational limitations, still hold for low-luminosity galaxies out to $z\simeq5$. With this in mind, we have discussed how our results depend on the adopted $L_\mathrm{IR}-L^\prime_\mathrm{CO}$ correlation. To compute the luminosities of CO lines with upper level $J\geq2$ from the CO(1--0) line luminosity, we rely on empirical CO spectral line energy distributions, which are a tracer of the excitation state of the gas. Since this is a particularly unknown for high-redshift low-luminosity galaxies, we discuss predicted CO line luminosities assuming two extreme CO SLEDs -- Milky Way and M82 -- which should bracket all possible CO excitations in these galaxies (assuming star formation is the only excitation source, i.e. no AGN).
Considering these uncertainties, our predictions are the best possible based on the current empirical knowledge of star-forming galaxies.

A possible caveat of the optical selection of our sample is that we may be missing very optically-faint dust-obscured sources, which would not be included in the \cite{Coe2006} optical catalog but could be bright in the (sub-)millimeter. This is most relevant for galaxies with $z>2$, where our optical catalogue becomes incomplete (Fig.~\ref{redshift}). However, based on how well our predicted extragalactic background light and number counts at 870\mic\ agree with observations, we conclude that such sources are probably not dominant. If very dust-obscured sources that are not included in our catalog exist in the considered field, our optically-based predictions can be considered a lower limit for the number of detections in a (sub-)millimeter survey.

Barring cosmic variance, our predictions should provide a statistical representation of an arbitrarily chosen region on the sky, and can be used to plan deep field observations in the (sub-)millimeter regime. 
We have illustrated the usefulness of these predictions by estimating the expected number of continuum and line detections in a 500 hour deep survey of the Hubble UDF with ALMA. For example, at 230~GHz (ALMA band 6), considering the required number of pointings necessary to cover the total area of the UDF (given the primary beam size), we expect a continuum sensitivity of $5.1~\mu$Jy (computed using the ALMA sensitivity calculator, assuming a bandwidth of 8~GHz). According to our predictions, this would yield a total of $\sim 600$ $3\sigma$-detections of galaxies distributed uniformly in redshift, with star formation rates down to 1~M$_\odot$~yr$^{-1}$ and infrared luminosities down to $10^{10}$~\lsun. We also discuss the predicted number of CO line detections when performing a scan over the whole frequency range covered by each band in steps of 8~GHz (the interval covered in each frequency setting). In 500 hours, the line sensitivity reached by such a frequency scan of each UDF pointing in band 6 is $84~\mu$Jy (assuming a 300~km~s$^{-1}$ bandwidth), which would yield between 19 and 137 detections, depending on the CO excitation.

The predictions presented in this paper will help to plan future (sub-)millimeter line and continuum deep field campaigns with new cutting-edge interferometers, and will also serve as a benchmark of our current empirical knowledge of dust continuum emission and CO/\cii\ line emission, to which these future deep observations can be compared.

%******************************************************************************

\section*{Acknowledgements}

We thank the anonymous referee for comments that helped improve the quality and clarity of this paper. We would also like to thank Vassilis Charmandaris and Eric Murphy for insightful comments on the manuscript.
EdC acknowledges funding through the ERC grant `Cosmic Dawn'.
RD acknowledges funding from Germany's national research center for aeronautics and space (DLR, project FKZ 50 OR 1104).
{\it Herschel} is an ESA space observatory with science instruments provided by European-led Principal Investigator consortia and with important participation from NASA.

%******************************************************************************

%******************************************************************************

% Bibliography and bibfile
\def\aj{AJ}
\def\araa{ARA\&A}
\def\apj{ApJ}
\def\apjl{ApJ}
\def\apjs{ApJS}
\def\apss{Ap\&SS}
\def\aap{A\&A}
\def\aapr{A\&A~Rev.}
\def\aaps{A\&AS}
\def\mnras{MNRAS}
\def\pasp{PASP}
\def\pasj{PASJ}
\def\qjras{QJRAS}
\def\nat{Nature}

\def\aplett{Astrophys.~Lett.}
\def\aas{AAS}
\let\astap=\aap
\let\apjlett=\apjl
\let\apjsupp=\apjs
\let\applopt=\ao

\bibliographystyle{apj}
%\bibliography{bib_dacunha}

\begin{thebibliography}{69}
\expandafter\ifx\csname natexlab\endcsname\relax\def\natexlab#1{#1}\fi

\bibitem[{{Bell}(2003)}]{Bell2003}
{Bell}, E.~F. 2003, \apj, 586, 794

\bibitem[{{Bell} {et~al.}(2003){Bell}, {McIntosh}, {Katz}, \&
  {Weinberg}}]{Bell2003b}
{Bell}, E.~F., {McIntosh}, D.~H., {Katz}, N., \& {Weinberg}, M.~D. 2003, \apjs,
  149, 289

\bibitem[{{Borys} {et~al.}(2003){Borys}, {Chapman}, {Halpern}, \&
  {Scott}}]{Borys2003}
{Borys}, C., {Chapman}, S., {Halpern}, M., \& {Scott}, D. 2003, \mnras, 344,
  385

\bibitem[{{Boselli} {et~al.}(2002){Boselli}, {Gavazzi}, {Lequeux}, \&
  {Pierini}}]{Boselli2002}
{Boselli}, A., {Gavazzi}, G., {Lequeux}, J., \& {Pierini}, D. 2002, \aap, 385,
  454

\bibitem[{{Bouwens} {et~al.}(2010){Bouwens}, {Illingworth}, {Oesch},
  {Stiavelli}, {van Dokkum}, {Trenti}, {Magee}, {Labb{\'e}}, {Franx},
  {Carollo}, \& {Gonzalez}}]{Bouwens2010}
{Bouwens}, R.~J., {Illingworth}, G.~D., {Oesch}, P.~A., {et~al.} 2010, \apjl,
  709, L133

\bibitem[{{Brauher} {et~al.}(2008){Brauher}, {Dale}, \& {Helou}}]{Brauher2008}
{Brauher}, J.~R., {Dale}, D.~A., \& {Helou}, G. 2008, \apjs, 178, 280

\bibitem[{{Bruzual} \& {Charlot}(2003)}]{Bruzual2003}
{Bruzual}, G., \& {Charlot}, S. 2003, \mnras, 344, 1000

\bibitem[{{Charlot} \& {Fall}(2000)}]{Charlot2000}
{Charlot}, S., \& {Fall}, S.~M. 2000, \apj, 539, 718

\bibitem[{{Coe} {et~al.}(2006){Coe}, {Ben{\'{\i}}tez}, {S{\'a}nchez}, {Jee},
  {Bouwens}, \& {Ford}}]{Coe2006}
{Coe}, D., {Ben{\'{\i}}tez}, N., {S{\'a}nchez}, S.~F., {et~al.} 2006, \aj, 132,
  926

\bibitem[{{Combes} {et~al.}(1999){Combes}, {Maoli}, \& {Omont}}]{Combes1999}
{Combes}, F., {Maoli}, R., \& {Omont}, A. 1999, \aap, 345, 369

\bibitem[{{Condon}(1992)}]{Condon1992}
{Condon}, J.~J. 1992, \araa, 30, 575

\bibitem[{{Coppin} {et~al.}(2006){Coppin}, {Chapin}, {Mortier}, \& {et
  al.}}]{Coppin2006}
{Coppin}, K., {Chapin}, E.~L., {Mortier}, A.~M.~J., \& {et al.} 2006, \mnras,
  372, 1621

\bibitem[{{Cox} {et~al.}(2011){Cox}, {Krips}, {Neri}, {Omont}, \& {et
  al.}}]{Cox2011}
{Cox}, P., {Krips}, M., {Neri}, R., {Omont}, A., \& {et al.} 2011, \apj, 740,
  63

\bibitem[{{da Cunha} {et~al.}(2008){da Cunha}, {Charlot}, \&
  {Elbaz}}]{daCunha2008}
{da Cunha}, E., {Charlot}, S., \& {Elbaz}, D. 2008, \mnras, 388, 1595

\bibitem[{{da Cunha} {et~al.}(2010){da Cunha}, {Charmandaris},
  {D{\'{\i}}az-Santos}, {Armus}, {Marshall}, \& {Elbaz}}]{daCunha2010b}
{da Cunha}, E., {Charmandaris}, V., {D{\'{\i}}az-Santos}, T., {et~al.} 2010,
  \aap, 523, A78+

\bibitem[{{Daddi} {et~al.}(2008){Daddi}, {Dannerbauer}, {Elbaz}, {Dickinson},
  {Morrison}, {Stern}, \& {Ravindranath}}]{Daddi2008}
{Daddi}, E., {Dannerbauer}, H., {Elbaz}, D., {et~al.} 2008, \apjl, 673, L21

\bibitem[{{Daddi} {et~al.}(2009{\natexlab{a}}){Daddi}, {Dannerbauer}, {Krips},
  {Walter}, {Dickinson}, {Elbaz}, \& {Morrison}}]{Daddi2009b}
{Daddi}, E., {Dannerbauer}, H., {Krips}, M., {et~al.} 2009{\natexlab{a}},
  \apjl, 695, L176

\bibitem[{{Daddi} {et~al.}(2005){Daddi}, {Dickinson}, {Chary}, {Pope},
  {Morrison}, {Alexander}, {Bauer}, {Brandt}, {Giavalisco}, {Ferguson}, {Lee},
  {Lehmer}, {Papovich}, \& {Renzini}}]{Daddi2005}
{Daddi}, E., {Dickinson}, M., {Chary}, R., {et~al.} 2005, \apjl, 631, L13

\bibitem[{{Daddi} {et~al.}(2007){Daddi}, {Dickinson}, {Morrison}, {Chary},
  {Cimatti}, {Elbaz}, {Frayer}, {Renzini}, {Pope}, {Alexander}, {Bauer},
  {Giavalisco}, {Huynh}, {Kurk}, \& {Mignoli}}]{Daddi2007}
{Daddi}, E., {Dickinson}, M., {Morrison}, G., {et~al.} 2007, \apj, 670, 156

\bibitem[{{Daddi} {et~al.}(2009{\natexlab{b}}){Daddi}, {Dannerbauer}, {Stern},
  {Dickinson}, {Morrison}, {Elbaz}, {Giavalisco}, {Mancini}, {Pope}, \&
  {Spinrad}}]{Daddi2009a}
{Daddi}, E., {Dannerbauer}, H., {Stern}, D., {et~al.} 2009{\natexlab{b}}, \apj,
  694, 1517

\bibitem[{{Daddi} {et~al.}(2010{\natexlab{a}}){Daddi}, {Elbaz}, {Walter},
  {Bournaud}, {Salmi}, {Carilli}, {Dannerbauer}, {Dickinson}, {Monaco}, \&
  {Riechers}}]{Daddi2010b}
{Daddi}, E., {Elbaz}, D., {Walter}, F., {et~al.} 2010{\natexlab{a}}, \apjl,
  714, L118

\bibitem[{{Daddi} {et~al.}(2010{\natexlab{b}}){Daddi}, {Bournaud}, {Walter},
  {Dannerbauer}, {Carilli}, {Dickinson}, {Elbaz}, {Morrison}, {Riechers},
  {Onodera}, {Salmi}, {Krips}, \& {Stern}}]{Daddi2010}
{Daddi}, E., {Bournaud}, F., {Walter}, F., {et~al.} 2010{\natexlab{b}}, \apj,
  713, 686

\bibitem[{{Dale} \& {Helou}(2002)}]{Dale2002}
{Dale}, D.~A., \& {Helou}, G. 2002, \apj, 576, 159

\bibitem[{{de Looze} {et~al.}(2011){de Looze}, {Baes}, {Bendo}, {Cortese}, \&
  {Fritz}}]{deLooze2011}
{de Looze}, I., {Baes}, M., {Bendo}, G.~J., {Cortese}, L., \& {Fritz}, J. 2011,
  \mnras, 416, 2712

\bibitem[{{Dunlop}(2011)}]{Dunlop2011}
{Dunlop}, J.~S. 2011, ArXiv e-prints

\bibitem[{{Elbaz} {et~al.}(2011){Elbaz}, {Dickinson}, {Hwang}, \& {et
  al.}}]{Elbaz2011}
{Elbaz}, D., {Dickinson}, M., {Hwang}, H.~S., \& {et al.} 2011, \aap, 533, A119

\bibitem[{{Fixsen} {et~al.}(1998){Fixsen}, {Dwek}, {Mather}, {Bennett}, \&
  {Shafer}}]{Fixsen1998}
{Fixsen}, D.~J., {Dwek}, E., {Mather}, J.~C., {Bennett}, C.~L., \& {Shafer},
  R.~A. 1998, \apj, 508, 123

\bibitem[{{Geach} {et~al.}(2011){Geach}, {Smail}, {Moran}, {MacArthur},
  {Lagos}, \& {Edge}}]{Geach2011}
{Geach}, J.~E., {Smail}, I., {Moran}, S.~M., {et~al.} 2011, \apjl, 730, L19

\bibitem[{{Genzel} {et~al.}(2010){Genzel}, {Tacconi}, {Gracia-Carpio},
  {Sternberg}, {Cooper}, {Shapiro}, {Bolatto}, {Bouch{\'e}}, {Bournaud},
  {Burkert}, {Combes}, {Comerford}, {Cox}, {Davis}, {Schreiber},
  {Garcia-Burillo}, {Lutz}, {Naab}, {Neri}, {Omont}, {Shapley}, \&
  {Weiner}}]{Genzel2010}
{Genzel}, R., {Tacconi}, L.~J., {Gracia-Carpio}, J., {et~al.} 2010, \mnras,
  407, 2091

\bibitem[{{Graci{\'a}-Carpio} {et~al.}(2011){Graci{\'a}-Carpio}, {Sturm},
  {Hailey-Dunsheath}, \& {et al.}}]{Gracia2011}
{Graci{\'a}-Carpio}, J., {Sturm}, E., {Hailey-Dunsheath}, S., \& {et al.} 2011,
  \apjl, 728, L7

\bibitem[{{Helou} {et~al.}(1985){Helou}, {Soifer}, \&
  {Rowan-Robinson}}]{Helou1985}
{Helou}, G., {Soifer}, B.~T., \& {Rowan-Robinson}, M. 1985, \apjl, 298, L7

\bibitem[{{Hopkins} \& {Beacom}(2006)}]{Hopkins2006}
{Hopkins}, A.~M., \& {Beacom}, J.~F. 2006, \apj, 651, 142

\bibitem[{{Ilbert} {et~al.}(2010){Ilbert}, {Salvato}, {Le Floc'h}, {Aussel},
  {Capak}, {McCracken}, {Mobasher}, {Kartaltepe}, {Scoville}, {Sanders},
  {Arnouts}, {Bundy}, {Cassata}, {Kneib}, {Koekemoer}, {Le F{\`e}vre}, {Lilly},
  {Surace}, {Taniguchi}, {Tasca}, {Thompson}, {Tresse}, {Zamojski}, {Zamorani},
  \& {Zucca}}]{Ilbert2010}
{Ilbert}, O., {Salvato}, M., {Le Floc'h}, E., {et~al.} 2010, \apj, 709, 644

\bibitem[{{Karim} {et~al.}(2011){Karim}, {Schinnerer},
  {Mart{\'{\i}}nez-Sansigre}, {Sargent}, {van der Wel}, {Rix}, {Ilbert},
  {Smol{\v c}i{\'c}}, {Carilli}, {Pannella}, {Koekemoer}, {Bell}, \&
  {Salvato}}]{Karim2011}
{Karim}, A., {Schinnerer}, E., {Mart{\'{\i}}nez-Sansigre}, A., {et~al.} 2011,
  \apj, 730, 61

\bibitem[{{Knudsen} {et~al.}(2008){Knudsen}, {van der Werf}, \&
  {Kneib}}]{Knudsen2008}
{Knudsen}, K.~K., {van der Werf}, P.~P., \& {Kneib}, J.-P. 2008, \mnras, 384,
  1611

\bibitem[{{Lagos} {et~al.}(2011){Lagos}, {Baugh}, {Lacey}, {Benson}, {Kim}, \&
  {Power}}]{Lagos2011}
{Lagos}, C.~D.~P., {Baugh}, C.~M., {Lacey}, C.~G., {et~al.} 2011, \mnras, 418,
  1649

\bibitem[{{Le F{\`e}vre} {et~al.}(2005){Le F{\`e}vre}, {Vettolani}, {Garilli},
  {Tresse}, {Bottini}, {Le Brun}, {Maccagni}, {Picat}, {Scaramella},
  {Scodeggio}, {Zanichelli}, {Adami}, {Arnaboldi}, {Arnouts}, {Bardelli},
  {Bolzonella}, {Cappi}, {Charlot}, {Ciliegi}, {Contini}, {Foucaud},
  {Franzetti}, {Gavignaud}, {Guzzo}, {Ilbert}, {Iovino}, {McCracken}, {Marano},
  {Marinoni}, {Mathez}, {Mazure}, {Meneux}, {Merighi}, {Paltani}, {Pell{\`o}},
  {Pollo}, {Pozzetti}, {Radovich}, {Zamorani}, {Zucca}, {Bondi}, {Bongiorno},
  {Busarello}, {Lamareille}, {Mellier}, {Merluzzi}, {Ripepi}, \&
  {Rizzo}}]{LeFevre2005}
{Le F{\`e}vre}, O., {Vettolani}, G., {Garilli}, B., {et~al.} 2005, \aap, 439,
  845

\bibitem[{{Lilly} {et~al.}(1996){Lilly}, {Le Fevre}, {Hammer}, \&
  {Crampton}}]{Lilly1996}
{Lilly}, S.~J., {Le Fevre}, O., {Hammer}, F., \& {Crampton}, D. 1996, \apjl,
  460, L1

\bibitem[{{Lilly} {et~al.}(2007){Lilly}, {Le F{\`e}vre}, {Renzini}, {Zamorani},
  {Scodeggio}, {Contini}, {Carollo}, {Hasinger}, {Kneib}, {Iovino}, {Le Brun},
  {Maier}, {Mainieri}, {Mignoli}, {Silverman}, {Tasca}, {Bolzonella},
  {Bongiorno}, {Bottini}, {Capak}, {Caputi}, {Cimatti}, {Cucciati}, {Daddi},
  {Feldmann}, {Franzetti}, {Garilli}, {Guzzo}, {Ilbert}, {Kampczyk}, {Kovac},
  {Lamareille}, {Leauthaud}, {Borgne}, {McCracken}, {Marinoni}, {Pello},
  {Ricciardelli}, {Scarlata}, {Vergani}, {Sanders}, {Schinnerer}, {Scoville},
  {Taniguchi}, {Arnouts}, {Aussel}, {Bardelli}, {Brusa}, {Cappi}, {Ciliegi},
  {Finoguenov}, {Foucaud}, {Franceschini}, {Halliday}, {Impey}, {Knobel},
  {Koekemoer}, {Kurk}, {Maccagni}, {Maddox}, {Marano}, {Marconi}, {Meneux},
  {Mobasher}, {Moreau}, {Peacock}, {Porciani}, {Pozzetti}, {Scaramella},
  {Schiminovich}, {Shopbell}, {Smail}, {Thompson}, {Tresse}, {Vettolani},
  {Zanichelli}, \& {Zucca}}]{Lilly2007}
{Lilly}, S.~J., {Le F{\`e}vre}, O., {Renzini}, A., {et~al.} 2007, \apjs, 172,
  70

\bibitem[{{Madau} {et~al.}(1996){Madau}, {Ferguson}, {Dickinson}, {Giavalisco},
  {Steidel}, \& {Fruchter}}]{Madau1996}
{Madau}, P., {Ferguson}, H.~C., {Dickinson}, M.~E., {et~al.} 1996, \mnras, 283,
  1388

\bibitem[{{Magdis} {et~al.}(2011){Magdis}, {Elbaz}, {Dickinson}, {Hwang},
  {Charmandaris}, {Armus}, {Daddi}, {Le Floc'h}, {Aussel}, {Dannerbauer},
  {Rigopoulou}, {Buat}, {Morrison}, {Mullaney}, {Lutz}, {Scott}, {Coia},
  {Pope}, {Pannella}, {Altieri}, {Burgarella}, {Bethermin}, {Dasyra},
  {Kartaltepe}, {Leiton}, {Magnelli}, {Popesso}, \& {Valtchanov}}]{Magdis2011}
{Magdis}, G.~E., {Elbaz}, D., {Dickinson}, M., {et~al.} 2011, \aap, 534, A15

\bibitem[{{Obreschkow} {et~al.}(2009{\natexlab{a}}){Obreschkow}, {Croton}, {De
  Lucia}, {Khochfar}, \& {Rawlings}}]{Obreschkow2009a}
{Obreschkow}, D., {Croton}, D., {De Lucia}, G., {Khochfar}, S., \& {Rawlings},
  S. 2009{\natexlab{a}}, \apj, 698, 1467

\bibitem[{{Obreschkow} {et~al.}(2009{\natexlab{b}}){Obreschkow}, {Heywood},
  {Kl{\"o}ckner}, \& {Rawlings}}]{Obreschkow2009}
{Obreschkow}, D., {Heywood}, I., {Kl{\"o}ckner}, H.-R., \& {Rawlings}, S.
  2009{\natexlab{b}}, \apj, 702, 1321

\bibitem[{{Obreschkow} {et~al.}(2011){Obreschkow}, {Heywood}, \&
  {Rawlings}}]{Obreschkow2011}
{Obreschkow}, D., {Heywood}, I., \& {Rawlings}, S. 2011, \apj, 743, 84

\bibitem[{{Oliver} {et~al.}(2010){Oliver}, {Wang}, {Smith}, {Altieri},
  {Amblard}, {Arumugam}, {Auld}, {Aussel}, {Babbedge}, {Blain}, {Bock},
  {Boselli}, {Buat}, {Burgarella}, {Castro-Rodr{\'{\i}}guez}, {Cava},
  {Chanial}, {Clements}, {Conley}, {Conversi}, {Cooray}, {Dowell}, {Dwek},
  {Eales}, {Elbaz}, {Fox}, {Franceschini}, {Gear}, {Glenn}, {Griffin},
  {Halpern}, {Hatziminaoglou}, {Ibar}, {Isaak}, {Ivison}, {Lagache},
  {Levenson}, {Lu}, {Madden}, {Maffei}, {Mainetti}, {Marchetti},
  {Mitchell-Wynne}, {Mortier}, {Nguyen}, {O'Halloran}, {Omont}, {Page},
  {Panuzzo}, {Papageorgiou}, {Pearson}, {P{\'e}rez-Fournon}, {Pohlen},
  {Rawlings}, {Raymond}, {Rigopoulou}, {Rizzo}, {Roseboom}, {Rowan-Robinson},
  {S{\'a}nchez Portal}, {Savage}, {Schulz}, {Scott}, {Seymour}, {Shupe},
  {Stevens}, {Symeonidis}, {Trichas}, {Tugwell}, {Vaccari}, {Valiante},
  {Valtchanov}, {Vieira}, {Vigroux}, {Ward}, {Wright}, {Xu}, \&
  {Zemcov}}]{Oliver2010}
{Oliver}, S.~J., {Wang}, L., {Smith}, A.~J., {et~al.} 2010, \aap, 518, L21

\bibitem[{{Power} {et~al.}(2010){Power}, {Baugh}, \& {Lacey}}]{Power2010}
{Power}, C., {Baugh}, C.~M., \& {Lacey}, C.~G. 2010, \mnras, 406, 43

\bibitem[{{Puget} {et~al.}(1996){Puget}, {Abergel}, {Bernard}, {Boulanger},
  {Burton}, {Desert}, \& {Hartmann}}]{Puget1996}
{Puget}, J.-L., {Abergel}, A., {Bernard}, J.-P., {et~al.} 1996, \aap, 308, L5

\bibitem[{{Reddy} {et~al.}(2006){Reddy}, {Steidel}, {Fadda}, {Yan}, {Pettini},
  {Shapley}, {Erb}, \& {Adelberger}}]{Reddy2006}
{Reddy}, N.~A., {Steidel}, C.~C., {Fadda}, D., {et~al.} 2006, \apj, 644, 792

\bibitem[{{Riechers} {et~al.}(2006){Riechers}, {Walter}, {Carilli}, {Knudsen},
  {Lo}, {Benford}, {Staguhn}, {Hunter}, {Bertoldi}, {Henkel}, {Menten},
  {Weiss}, {Yun}, \& {Scoville}}]{Riechers2006}
{Riechers}, D.~A., {Walter}, F., {Carilli}, C.~L., {et~al.} 2006, \apj, 650,
  604

\bibitem[{{Rodighiero} {et~al.}(2011){Rodighiero}, {Daddi}, {Baronchelli}, \&
  {et al.}}]{Rodighiero2011}
{Rodighiero}, G., {Daddi}, E., {Baronchelli}, I., \& {et al.} 2011, \apjl, 739,
  L40

\bibitem[{{Sargent} {et~al.}(2012){Sargent}, {B{\'e}thermin}, {Daddi}, \&
  {Elbaz}}]{Sargent2012}
{Sargent}, M.~T., {B{\'e}thermin}, M., {Daddi}, E., \& {Elbaz}, D. 2012, \apjl,
  747, L31

\bibitem[{{Sargent} {et~al.}(2010){Sargent}, {Schinnerer}, {Murphy}, {Carilli},
  {Helou}, {Aussel}, {Le Floc'h}, {Frayer}, {Ilbert}, {Oesch}, {Salvato},
  {Smol{\v c}i{\'c}}, {Kartaltepe}, \& {Sanders}}]{Sargent2010}
{Sargent}, M.~T., {Schinnerer}, E., {Murphy}, E., {et~al.} 2010, \apjl, 714,
  L190

\bibitem[{{Scott} {et~al.}(2006){Scott}, {Dunlop}, \& {Serjeant}}]{Scott2006}
{Scott}, S.~E., {Dunlop}, J.~S., \& {Serjeant}, S. 2006, \mnras, 370, 1057

\bibitem[{{Scott} {et~al.}(2002){Scott}, {Fox}, {Dunlop}, {Serjeant},
  {Peacock}, {Ivison}, {Oliver}, {Mann}, {Lawrence}, {Efstathiou},
  {Rowan-Robinson}, {Hughes}, {Archibald}, {Blain}, \& {Longair}}]{Scott2002}
{Scott}, S.~E., {Fox}, M.~J., {Dunlop}, J.~S., {et~al.} 2002, \mnras, 331, 817

\bibitem[{{Silva} {et~al.}(1998){Silva}, {Granato}, {Bressan}, \&
  {Danese}}]{Silva1998}
{Silva}, L., {Granato}, G.~L., {Bressan}, A., \& {Danese}, L. 1998, \apj, 509,
  103

\bibitem[{{Smail} {et~al.}(2002){Smail}, {Ivison}, {Blain}, \&
  {Kneib}}]{Smail2002}
{Smail}, I., {Ivison}, R.~J., {Blain}, A.~W., \& {Kneib}, J.-P. 2002, \mnras,
  331, 495

\bibitem[{{Smol{\v c}i{\'c}} {et~al.}(2009){Smol{\v c}i{\'c}}, {Schinnerer},
  {Zamorani}, {Bell}, {Bondi}, {Carilli}, {Ciliegi}, {Mobasher}, {Paglione},
  {Scodeggio}, \& {Scoville}}]{Smolcic2009}
{Smol{\v c}i{\'c}}, V., {Schinnerer}, E., {Zamorani}, G., {et~al.} 2009, \apj,
  690, 610

\bibitem[{{Solomon} \& {Vanden Bout}(2005)}]{Solomon2005}
{Solomon}, P.~M., \& {Vanden Bout}, P.~A. 2005, \araa, 43, 677

\bibitem[{{Springel} {et~al.}(2005){Springel}, {White}, {Jenkins}, {Frenk},
  {Yoshida}, {Gao}, {Navarro}, {Thacker}, {Croton}, {Helly}, {Peacock}, {Cole},
  {Thomas}, {Couchman}, {Evrard}, {Colberg}, \& {Pearce}}]{Springel2005}
{Springel}, V., {White}, S.~D.~M., {Jenkins}, A., {et~al.} 2005, \nat, 435, 629

\bibitem[{{Stacey} {et~al.}(1991){Stacey}, {Geis}, {Genzel}, {Lugten},
  {Poglitsch}, {Sternberg}, \& {Townes}}]{Stacey1991}
{Stacey}, G.~J., {Geis}, N., {Genzel}, R., {et~al.} 1991, \apj, 373, 423

\bibitem[{{Stacey} {et~al.}(2010){Stacey}, {Hailey-Dunsheath}, {Ferkinhoff},
  {Nikola}, {Parshley}, {Benford}, {Staguhn}, \& {Fiolet}}]{Stacey2010}
{Stacey}, G.~J., {Hailey-Dunsheath}, S., {Ferkinhoff}, C., {et~al.} 2010, \apj,
  724, 957

\bibitem[{{Steidel} {et~al.}(2003){Steidel}, {Adelberger}, {Shapley},
  {Pettini}, {Dickinson}, \& {Giavalisco}}]{Steidel2003}
{Steidel}, C.~C., {Adelberger}, K.~L., {Shapley}, A.~E., {et~al.} 2003, \apj,
  592, 728

\bibitem[{{Tacconi} {et~al.}(2010){Tacconi}, {Genzel}, {Neri}, {Cox}, {Cooper},
  {Shapiro}, {Bolatto}, {Bouch{\'e}}, {Bournaud}, {Burkert}, {Combes},
  {Comerford}, {Davis}, {Schreiber}, {Garcia-Burillo}, {Gracia-Carpio}, {Lutz},
  {Naab}, {Omont}, {Shapley}, {Sternberg}, \& {Weiner}}]{Tacconi2010}
{Tacconi}, L.~J., {Genzel}, R., {Neri}, R., {et~al.} 2010, \nat, 463, 781

\bibitem[{{Walter} {et~al.}(2011){Walter}, {Carilli}, \& {Daddi}}]{Walter2011}
{Walter}, F., {Carilli}, C., \& {Daddi}, E. 2011, ArXiv e-prints

\bibitem[{{Walter} {et~al.}(2012){Walter}, {Decarli}, {Carilli}, {Bertoldi},
  {Cox}, {da Cunha}, {Daddi}, {Dickinson}, {Downes}, {Elbaz}, {Ellis}, {Hodge},
  {Neri}, {Riechers}, {Weiss}, {Bell}, {Dannerbauer}, {Krips}, {Krumholz},
  {Lentati}, {Maiolino}, {Menten}, {Rix}, {Robertson}, {Spinrad}, {Stark}, \&
  {Stern}}]{Walter2012}
{Walter}, F., {Decarli}, R., {Carilli}, C., {et~al.} 2012, \nat, 486, 233

\bibitem[{{Weiss} {et~al.}(2007){Weiss}, {Downes}, {Walter}, \&
  {Henkel}}]{Weiss2007}
{Weiss}, A., {Downes}, D., {Walter}, F., \& {Henkel}, C. 2007, in Astronomical
  Society of the Pacific Conference Series, Vol. 375, From Z-Machines to ALMA:
  (Sub)Millimeter Spectroscopy of Galaxies, ed. {A.~J.~Baker, J.~Glenn,
  A.~I.~Harris, J.~G.~Mangum, \& M.~S.~Yun }, 25--+

\bibitem[{{Wei{\ss}} {et~al.}(2009){Wei{\ss}}, {Kov{\'a}cs}, {Coppin}, \& {et
  al.}}]{Weiss2009}
{Wei{\ss}}, A., {Kov{\'a}cs}, A., {Coppin}, K., \& {et al.} 2009, \apj, 707,
  1201

\bibitem[{{Wuyts} {et~al.}(2008){Wuyts}, {Labb{\'e}}, {Schreiber}, {Franx},
  {Rudnick}, {Brammer}, \& {van Dokkum}}]{Wuyts2008}
{Wuyts}, S., {Labb{\'e}}, I., {Schreiber}, N.~M.~F., {et~al.} 2008, \apj, 682,
  985

\bibitem[{{Yun} {et~al.}(2001){Yun}, {Reddy}, \& {Condon}}]{Yun2001}
{Yun}, M.~S., {Reddy}, N.~A., \& {Condon}, J.~J. 2001, \apj, 554, 803

\end{thebibliography}

\appendix

\section{Comparison with other SED templates}
\label{appendix_seds}

\begin{figure*}
\begin{center}
\includegraphics[width=0.6\textwidth]{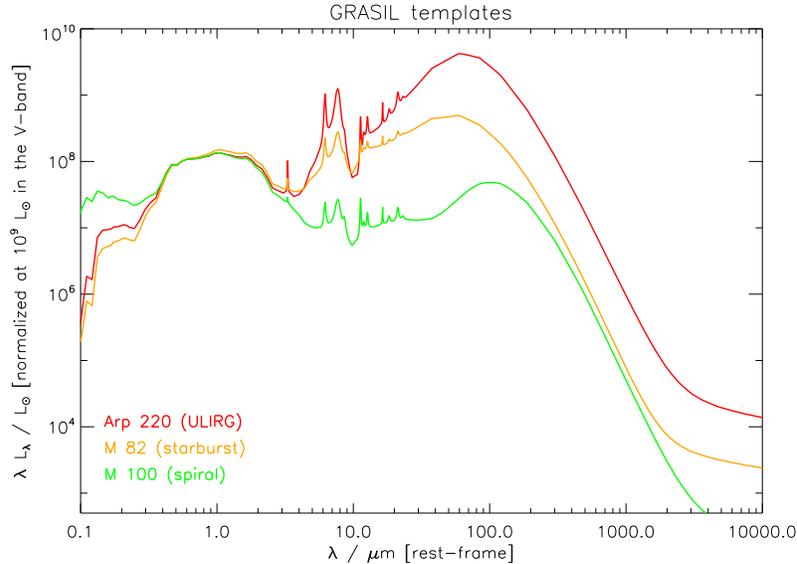}
\caption{Ultraviolet to millimetre spectral energy distributions of Arp\,220 (ULIRG), M\,82 (low-luminosity starburst), and M\,100 (spiral galaxy) from \protect\cite{Silva1998}, normalized in the $V$-band to emphasize their different optical-to-infrared ratios.}
\label{templates}
\end{center}
\end{figure*}

For reference, we compare our predicted flux densities with those obtained using three typical local galaxy SEDs: an ultra-luminous infrared galaxy (ULIRG), Arp\,220, a low-luminosity starburst, M\,82, and a typical spiral galaxy, M\,100. We use template SEDs from \cite{Silva1998} which cover the full spectrum from ultraviolet to radio wavelengths. In Fig.~\ref{templates}, we plot these SEDs normalized in the $V$-band to illustrate the large range in optical-to-infrared ratio spanned by these templates.

\begin{figure*}
\begin{center}
\includegraphics[width=0.7\textwidth]{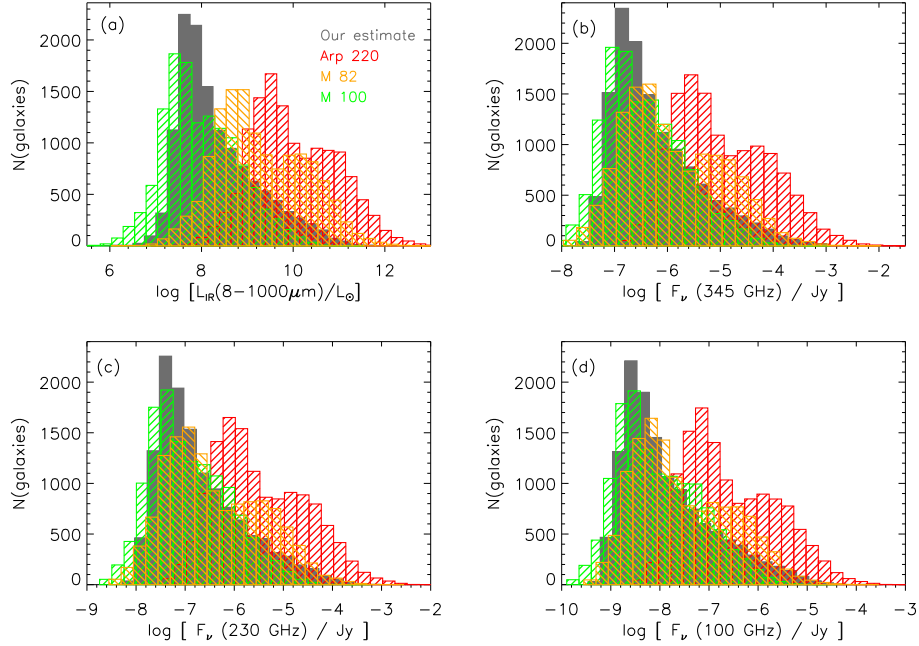}
\caption{Comparison between our estimates (using the \protect\citealt{daCunha2008} models) and estimates using 3 typical galaxy templates of (a) infrared luminosity; (b), (c) and (d) distribution of continuum flux density at 345, 230 and 100~GHz (ALMA bands 7, 6, 3; PdBI/NOEMA bands 4, 3, 1), respectively.}
\label{compare_templates}
\end{center}
\end{figure*}

We k-correct and re-scale each template to fit the observed optical magnitudes of each galaxy in our sample, and from that we predict what the continuum fluxes in the ALMA bands would be if the observed galaxy SED was the same as the template.

In Fig.~\ref{compare_templates}, we compare the distribution of the infrared luminosity of the galaxies and the continuum flux in three ALMA bands when using our \cite{daCunha2008} SED fits and when using these three templates. These distributions show the range of infrared luminosities and (sub-)mm fluxes of our observed galaxies if they all had optical-to-infrared ratios and infrared SEDs similar to Arp\,220, M\,82 and M\,100, versus our predictions, which make an educated guess by comparing the observed UV/optical SED of each galaxy with a library of models with a wide range of optical-to-infrared ratios and infrared SED shapes. It is clear that, if we assumed that all observed galaxies had Arp\,220-like SEDs, the typical infrared luminosities and predicted continuum fluxes would be typically 2 orders of magnitude higher. An M\,82 template, with a similar infrared SED shape but higher optical-to-infrared ratio, still produces infrared luminosities about one order of magnitude higher than our \cite{daCunha2008} SED fits, but the typical (sub-)mm continuum fluxes are similar to ours, except for part of the sample presenting an excess of fluxes compared to our estimates. The typical spiral template, M\,100, has the higher optical-to-infrared ratios and cooler typical dust temperature, and predicts similar, if not slightly lower, typical infrared luminosities and sub-mm continuum fluxes than our fits.

As in Section~\ref{sect:ebl}, we can compare the integrated continuum emission over the whole UDF, which, divided by the area of the field, allows us to get an estimate of the extragalactic background light (EBL). The predicted continuum fluxes at 345~GHz (ALMA 7) from our \cite{daCunha2008} SED fits yield a EBL of 45.6~Jy~deg$^{-2}$, which is consistent with observations, as discussed in Section~\ref{sect:ebl}. When using fixed SED templates, we get  389~Jy~deg$^{-2}$ (Arp\,220),  52.5~Jy~deg$^{-2}$ (M\,82) and  9.32~Jy~deg$^{-2}$ (M\,100), i.e. in the case of Arp\,220 we get too much EBL at 345~GHz by almost a factor of 10, while in the case of M\,100 we tend to underestimate the EBL at 345~GHz by about a factor of 5. 

\section{Redshift ranges where CO and \cii\ are observable in typical (sub-)mm bands}
\label{appendix_co_lines}

\begin{figure*}
\begin{center}
\includegraphics[width=0.7\textwidth]{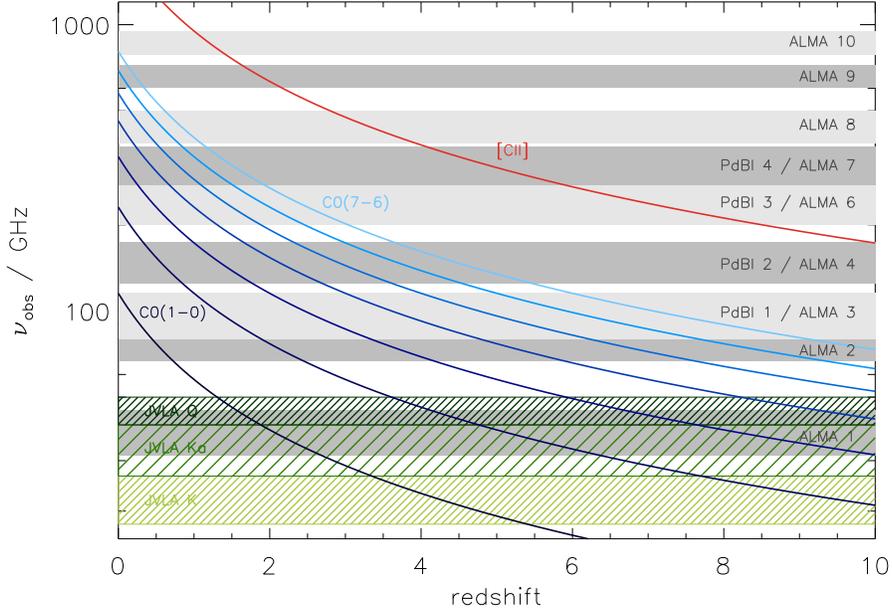}
\caption{Observed frequency of the CO lines CO(1--0) to CO(7--6) (from darker blue to lighter blue lines) and [C{\sc ii}] 158\mic\ (red line) as a function of redshift. The gray and green shaded
areas show the frequency ranges sampled by the ALMA, JVLA and PdBI/NOEMA bands considered in this paper.}
\label{co_detect}
\end{center}
\end{figure*}

For reference, we show in this appendix the redshift ranges with the CO lines (1--0) to (7--6) and the \cii\ line are observable in the typical (sub-)mm bands from ALMA, JVLA and PdBI/NOEMA (Fig.~\ref{co_detect} and Table~\ref{lines_z}).

\begin{sidewaystable}
\begin{center}
\caption{Redshift ranges where the CO and \cii\ lines are observable as a function of band frequency range.}
\begin{tabular}{l c c c c c c c c c}
\tableline
\tableline
& & {\bf CO(1--0)} & {\bf CO(2--1)} & {\bf CO(3--2)} & {\bf CO(4--3)} & {\bf CO(5--4)} & {\bf CO(6--5)} & {\bf CO(7--6)} & {\bf \protect[C {\sc ii\protect]}} \\
Band & Frequency range & 115.27 GHz & 230.54 GHz & 345.80 GHz & 461.04 GHz & 576.27 GHz & 691.47 GHz &806.65 GHz &1900.54 GHz \\
& / GHz &  &  & &  &  & & & \\
\tableline
JVLA K & 18 -- 26.5 & 3.35 -- 5.40 & 7.70 -- 11.8 & -- & -- & -- & -- & -- & -- \\
JVLA Ka & 26.5 -- 40 & 1.88 -- 3.35 & 4.76 -- 7.70 & 7.65 -- 12.0 & -- & -- & -- & -- & -- \\
ALMA 1 & 31.3 -- 45 & 1.56 -- 2.68 & 4.12 -- 6.37 & 6.68 -- 10.0 & 9.24 -- 13.7 & -- & -- & -- & -- \\
JVLA Q & 40 -- 50 & 1.31 -- 1.88 & 3.61 -- 4.76 & 5.91 -- 7.65 & 8.22 -- 10.5 & -- & -- & -- & -- \\
ALMA 2 & 67 -- 90 & 0.28 -- 0.72 & 1.56 -- 2.44 & 2.84 -- 4.16 & 4.12 -- 5.88 & 5.40 -- 7.60 & 6.68 -- 9.32 & 7.97 -- 11.0 & -- \\
ALMA 3 & 84 -- 116 & 0.00 -- 0.37 & 0.99 -- 1.74 & 1.98 -- 3.12 & 2.97 -- 4.49 & 3.97 -- 5.86 & 4.96 -- 7.23 & 5.95 -- 8.60 & -- \\
PdBI/NOEMA 1 & 80 -- 116 & 0.00 -- 0.44 & 0.99 -- 1.88 & 1.98 -- 3.32 & 2.97 -- 4.76 & 3.97 -- 6.20 & 4.96 -- 7.64 & 5.95 -- 9.08 & -- \\
ALMA 4 & 125 -- 163 & -- & 0.41 -- 0.84 & 1.12 -- 1.77 & 1.83 -- 2.69 & 2.54 -- 3.61 & 3.24 -- 4.53 & 3.95 -- 5.45 & -- \\
PdBI/NOEMA 2   & 129 -- 174 & -- & 0.32 -- 0.79 & 0.99 -- 1.68 & 1.66 -- 2.57 & 2.31 -- 3.47 & 2.97 -- 4.36 & 3.64 -- 5.25 & -- \\
ALMA 6 & 211 -- 275 & -- & 0.00 -- 0.09 & 0.26 -- 0.64 & 0.68 -- 1.19 & 1.10 -- 1.73 & 1.51 -- 2.28 & 1.93 -- 2.82 & 5.91 -- 8.01 \\
PdBI/NOEMA 3   & 201 -- 267 & -- & 0.00 -- 0.15 & 0.30 -- 0.72 & 0.73 -- 1.29 & 1.16 -- 1.87 & 1.59 -- 2.44 & 2.02 -- 3.01 & 6.12 -- 8.46 \\
ALMA 7 & 275 -- 373 & -- & -- & 0.00 -- 0.26 & 0.24 -- 0.68 & 0.55 -- 1.10 & 0.85 -- 1.51 & 1.16 -- 1.93 & 4.10 -- 5.91 \\
PdBI/NOEMA 4   & 277 -- 371 & -- & -- & 0.00 -- 0.25 & 0.24 -- 0.66 & 0.55 -- 1.08 & 0.86 -- 1.50 & 1.17 -- 1.91 & 4.12 -- 5.86 \\
ALMA 8 & 385 -- 500 & -- & -- & -- & 0.00 -- 0.20 & 0.15 -- 0.50 & 0.38 -- 0.80 & 0.61 -- 1.10 & 2.80 -- 3.94 \\
ALMA 9 & 602 -- 710 & -- & -- & -- & -- & -- & 0.00 -- 0.15 & 0.12 -- 0.34 & 1.68 -- 2.16 \\
ALMA 10 & 787 -- 950 & -- & -- & -- & -- & -- & -- & 0.00 -- 0.03 & 1.00 -- 1.41 \\
\tableline
\end{tabular}
\label{lines_z}
\end{center}
\end{sidewaystable}

\section{Predicted (sub-)mm fluxes for each galaxy in our sample}
\label{appendix_fluxes}

In this Appendix, we provide the predicted continuum and line fluxes for each galaxy in our sample. In Table~\ref{cont_predictions_table}, we list the predicted continuum fluxes (including confidence ranges) in the various 
(sub)mm bands. In Table~\ref{line_predictions_table}, we provide the predicted fluxes of the CO lines (using both Milky Way and M82 SLEDs) and of the \cii\ line.

\begin{sidewaystable}
\begin{center}
\caption{Predicted continuum fluxes and confidence ranges for the first 10 galaxies in our sample. The predicted fluxes are the median of the likelihood distribution
computed as described in Section~\ref{sect_method}, and the confidence ranges correspond to the 16th -- 84th percentile of the likelihood distribution.}
\begin{tabular}{l c c c c c c c c c c c c }
\tableline
\tableline
ID& R.A.& Dec & $z_\mathrm{phot}$ & $\log(F_\nu/\mathrm{Jy})$ & $\log(F_\nu/\mathrm{Jy})$ & $\log(F_\nu/\mathrm{Jy})$ & $\log(F_\nu/\mathrm{Jy})$ & $\log(F_\nu/\mathrm{Jy})$ & $\log(F_\nu/\mathrm{Jy})$ & $\log(F_\nu/\mathrm{Jy})$ & $\log(F_\nu/\mathrm{Jy})$ & $\log(F_\nu/\mathrm{Jy})$ \\
  & & & & 38~GHz & 80~GHz & 100~GHz & 144~GHz & 230~GHz & 345~GHz & 430~GHz & 660~GHz & 870~GHz \\
\tableline
1 & 03:32:39.72 & -27:49:42.53 & 0.472&$-6.82^{+0.46}_{-0.33}$&$-6.59^{+0.50}_{-0.40}$ & $-6.36^{+0.52}_{-0.45}$& $-5.89^{+0.53}_{-0.50}$&$-5.16^{+0.53}_{-0.53}$ & $-4.57^{+0.53}_{-0.54}$ & $-4.34^{+0.54}_{-0.54}$ & $-3.79^{+0.54}_{-0.53}$ & $-3.57^{+0.53}_{-0.52}$ \\ 
2 & 03:32:39.48 & -27:49:45.42 & 2.703&$-8.56^{+0.37}_{-0.51}$&$-7.59^{+0.41}_{-0.51}$ & $-7.26^{+0.41}_{-0.51}$& $-6.80^{+0.41}_{-0.51}$&$-6.18^{+0.41}_{-0.51}$ & $-5.80^{+0.40}_{-0.51}$ & $-5.68^{+0.40}_{-0.50}$ & $-5.57^{+0.35}_{-0.51}$ & $-5.61^{+0.32}_{-0.50}$ \\ 
3 & 03:32:39.17 & -27:49:45.20 & 1.281&$-9.69^{+0.41}_{-0.51}$&$-9.12^{+0.49}_{-0.51}$ & $-8.82^{+0.51}_{-0.51}$& $-8.35^{+0.53}_{-0.50}$&$-7.64^{+0.52}_{-0.50}$ & $-7.12^{+0.53}_{-0.50}$ & $-6.93^{+0.52}_{-0.50}$ & $-6.53^{+0.50}_{-0.51}$ & $-6.41^{+0.45}_{-0.52}$ \\ 
4 & 03:32:39.11 & -27:49:44.90 & 3.791&$-8.32^{+0.43}_{-0.48}$&$-7.34^{+0.46}_{-0.52}$ & $-7.03^{+0.47}_{-0.51}$& $-6.57^{+0.47}_{-0.51}$&$-6.03^{+0.45}_{-0.50}$ & $-5.72^{+0.42}_{-0.48}$ & $-5.64^{+0.39}_{-0.48}$ & $-5.61^{+0.31}_{-0.45}$ & $-5.68^{+0.30}_{-0.45}$ \\ 
5 & 03:32:39.37 & -27:49:44.01 & 0.452&$-8.24^{+0.35}_{-0.36}$&$-8.02^{+0.36}_{-0.40}$ & $-7.80^{+0.36}_{-0.43}$& $-7.34^{+0.37}_{-0.46}$&$-6.59^{+0.37}_{-0.49}$ & $-6.01^{+0.38}_{-0.49}$ & $-5.78^{+0.38}_{-0.49}$ & $-5.22^{+0.38}_{-0.48}$ & $-5.01^{+0.38}_{-0.48}$ \\ 
6 & 03:32:39.43 & -27:49:44.04 & 2.769&$-8.90^{+0.47}_{-0.45}$&$-7.91^{+0.45}_{-0.48}$ & $-7.59^{+0.46}_{-0.48}$& $-7.11^{+0.46}_{-0.47}$&$-6.49^{+0.45}_{-0.48}$ & $-6.12^{+0.45}_{-0.48}$ & $-6.03^{+0.45}_{-0.48}$ & $-5.95^{+0.48}_{-0.43}$ & $-5.99^{+0.45}_{-0.38}$ \\ 
7 & 03:32:39.45 & -27:49:42.95 & 1.653&$-7.84^{+0.46}_{-0.33}$&$-7.07^{+0.46}_{-0.44}$ & $-6.74^{+0.46}_{-0.46}$& $-6.24^{+0.46}_{-0.47}$&$-5.55^{+0.46}_{-0.48}$ & $-5.07^{+0.45}_{-0.47}$ & $-4.91^{+0.46}_{-0.46}$ & $-4.59^{+0.46}_{-0.42}$ & $-4.55^{+0.44}_{-0.38}$ \\ 
8 & 03:32:39.54 & -27:49:28.35 & 0.636&$-6.01^{+0.94}_{-0.50}$&$-5.49^{+0.87}_{-0.58}$ & $-5.16^{+0.85}_{-0.63}$& $-4.66^{+0.85}_{-0.66}$&$-3.89^{+0.84}_{-0.67}$ & $-3.33^{+0.85}_{-0.66}$ & $-3.11^{+0.85}_{-0.65}$ & $-2.58^{+0.84}_{-0.61}$ & $-2.40^{+0.85}_{-0.59}$ \\ 
9 & 03:32:39.10 & -27:49:43.91 & 0.695&$-8.85^{+0.70}_{-0.43}$&$-8.45^{+0.73}_{-0.52}$ & $-8.18^{+0.74}_{-0.55}$& $-7.68^{+0.74}_{-0.58}$&$-6.95^{+0.75}_{-0.59}$ & $-6.38^{+0.74}_{-0.60}$ & $-6.16^{+0.74}_{-0.60}$ & $-5.66^{+0.75}_{-0.58}$ & $-5.49^{+0.75}_{-0.56}$ \\ 
10 & 03:32:39.52 & -27:49:43.20 & 1.029&$-8.62^{+0.71}_{-0.10}$&$-8.28^{+0.30}_{-0.71}$ & $-8.06^{+0.35}_{-0.66}$& $-7.62^{+0.41}_{-0.63}$&$-6.93^{+0.44}_{-0.60}$ & $-6.39^{+0.44}_{-0.60}$ & $-6.18^{+0.43}_{-0.60}$ & $-5.68^{+0.37}_{-0.65}$ & $-5.49^{+0.32}_{-0.68}$ \\ 
\tableline
\end{tabular}
\tablecomments{IDs are the same as in the \cite{Coe2006} catalog. This table is published in its entirety in the electronic edition the journal; a portion is shown here for guidance regarding its form and content.}
\label{cont_predictions_table}
\end{center}
\end{sidewaystable}

\begin{sidewaystable}
\begin{center}
\caption{Predicted CO and [CII] line fluxes for the first 10 galaxies in our sample. For the CO lines, we include the flux prediction using
a Milky Way CO SLED (MW) and a M82 center SLED (M82), as described in Section~\ref{sect:co}.}
\begin{tabular}{l c c c c c c c c c }
\tableline
\tableline
 ID & $z_\mathrm{phot}$ & $\log(S_\nu/\mathrm{Jy})$ & $\log(S_\nu/\mathrm{Jy})$ & $\log(S_\nu/\mathrm{Jy})$ & $\log(S_\nu/\mathrm{Jy})$ & $\log(S_\nu/\mathrm{Jy})$ & $\log(S_\nu/\mathrm{Jy})$ & $\log(S_\nu/\mathrm{Jy})$ & $\log(S_\nu/\mathrm{Jy})$ \\
  & & CO(1--0) & CO(2--1) & CO(3--2) & CO(4--3) & CO(5--4) & CO(6--5) & CO(7--6) & [CII] \\
  & &  & MW / M82 & MW / M82 & MW / M82 & MW / M82 & MW / M82 & MW / M82 & \\
\tableline
1 &  $0.472$&$-4.58$&$-4.28$$ / $$-3.99$ & $-4.19$ / $-3.66$& $-4.14$ / $-3.45$&$-4.27$ / $-3.31$ & $-4.50$ / $-3.24$ & $-4.63$ / $-3.30$ &  $-2.36$ \\ 
2 &  $2.703$&$-6.64$&$-6.34$ / $-6.04$ & $-6.25$ / $-5.72$& $-6.20$ / $-5.50$&$-6.32$ / $-5.37$ & $-6.56$ / $-5.30$ & $-6.69$ / $-5.36$ &  $-4.50$ \\ 
3 &  $1.281$&$-7.23$&$-6.93$ / $-6.64$ & $-6.84$ / $-6.31$& $-6.79$ / $-6.10$&$-6.92$ / $-5.96$ & $-7.16$ / $-5.90$ & $-7.28$ / $-5.96$ &  $-5.26$ \\ 
4 &  $3.791$&$-6.66$&$-6.35$ / $-6.06$ & $-6.26$ / $-5.73$& $-6.21$ / $-5.52$&$-6.34$ / $-5.38$ & $-6.58$ / $-5.32$ & $-6.70$ / $-5.38$ &  $-4.49$ \\ 
5 &  $0.452$&$-5.83$&$-5.52$ / $-5.23$ & $-5.44$ / $-4.91$& $-5.38$ / $-4.69$&$-5.51$ / $-4.55$ & $-5.75$ / $-4.49$ & $-5.88$ / $-4.55$ & $-3.78$ \\ 
6 &  $2.769$&$-6.95$&$-6.65$ / $-6.36$ & $-6.56$ / $-6.03$& $-6.51$ / $-5.82$&$-6.64$ / $-5.68$ & $-6.87$ / $-5.62$ & $-7.00$ / $-5.68$ &  $-4.86$ \\ 
7 &  $1.653$&$-5.73$&$-5.42$ / $-5.13$ & $-5.33$ / $-4.80$& $-5.28$ / $-4.59$&$-5.41$ / $-4.45$ & $-5.65$ / $-4.39$ & $-5.77$ / $-4.45$ &  $-3.52$ \\ 
8 &  $0.636$&$-3.94$&$-3.64$ / $-3.35$ & $-3.55$ / $-3.02$& $-3.50$ / $-2.81$&$-3.63$ / $-2.67$ & $-3.86$ / $-2.61$ & $-3.99$/ $-2.66$ &  $-1.60$ \\ 
9 &  $0.695$&$-6.41$&$-6.11$ / $-5.82$ & $-6.02$ / $-5.49$& $-5.97$ / $-5.28$&$-6.10$ / $-5.14$ & $-6.33$ / $-5.08$ & $-6.46$ / $-5.14$ &  $-4.40$ \\ 
10& $1.029$&$-6.27$&$-5.96$ / $-5.67$ & $-5.88$ / $-5.35$& $-5.82$ / $-5.52$&$-5.95$ / $-4.99$ & $-6.19$ / $-4.93$ & $-6.31$ / $-3.98$ &  $-4.18$ \\ 
\tableline
\end{tabular}
\tablecomments{IDs are the same as in the \cite{Coe2006} catalog. This table is published in its entirety in the electronic edition the journal; a portion is shown here for guidance regarding its form and content.}
\label{line_predictions_table}
\end{center}
\end{sidewaystable}

\end{document}